\documentclass[12pt]{article}
\usepackage{amsfonts,amsmath,amssymb}
\usepackage{scalerel}[2016/12/29]
\usepackage{graphicx}
\usepackage{tikz}
\usepackage{braket}

\usetikzlibrary{arrows}
\usetikzlibrary{decorations.markings}

\numberwithin{equation}{section}

\newcommand{\za}{{z_i}}

\newcommand{\T}{{\bar t}\hskip 0.02cm}
\newcommand{\s}{{\bar s}}

\def\be{\begin{equation}}
\def\ee{\end{equation}}
\def\bea{\begin{eqnarray}}
\def\eea{\end{eqnarray}}

\def\half{{\textstyle {1\over 2}}}

\def\bb{\hskip -0.5mm}
\def\hb{\hskip -0.25mm}

\usepackage{amsmath}
\usepackage{epsfig,graphicx}
\usepackage{graphicx}
\usepackage{tikz}

\def\be{\begin{equation}}
\def\ee{\end{equation}}
\def\bea{\begin{eqnarray}}
\def\eea{\end{eqnarray}}

\def\Dra#1{\left<\hskip -0.9mm\left< #1 \right.\right|}
\def\Cet#1{\left|\left. #1\right>\hskip -0.9mm\right>}

\def\tF{\hat{F}}
\def\tG{\hat{G}}

\def\tH{\hat{H}}

\begin{document}

\thispagestyle{plain}

\title{\bf\Large Length and area generating functions for height-restricted Motzkin meanders}

\author{Alexios P. Polychronakos}


\maketitle

%

{\em
{\centerline{Physics Department, the City College of New York, NY 10031, USA}}
\vskip .2 cm \centerline{and}
\vskip .2 cm
{\centerline{The Graduate Center, CUNY, New York, NY 10016, USA}
\vskip .2 cm
\centerline{\it apolychronakos@ccny.cuny.edu}
}}

\maketitle

\begin{abstract}
We derive the length and area generating function of planar height-restricted forward-moving discrete paths of increments
$\pm1$ or $0$ with arbitrary starting and ending points, the so-called Motzkin meanders, and the more general
length-area generating functions for Motzkin paths with markers monitoring the number of passages
from the two height boundaries (`floor' and `ceiling') and the time spent there. The results are obtained by
embedding Motzkin paths in a two-step anisotropic Dyck path process and using propagator, exclusion statistics
and bosonization techniques. We also present a cluster expansion of the
logarithm of the generating functions that makes their polynomial structure explicit. These results are relevant to
the derivation of statistical mechanical properties of physical systems such as polymers, vesicles, and solid-on-solid
interfaces.

\end{abstract}

\vskip 1cm

\vfill
\eject

\tableofcontents

\section{Introduction}

Random walks of given length and area on planar lattices are of inherent mathematical and physical interest. In mathematics, their combinatorial properties, statistics and generating functions are the subject of intense study. In physics, they arise either in actual diffusion processes near boundaries or, indirectly, in the quantum mechanics of particles moving in a periodic
two-dimensional potential. The Hofstadter problem is the canonical example of the latter, leading to the famous
``butterfly'' energy spectrum \cite{Hofstadter}.

In physical contexts, random walks are generated through the action of a Hamiltonian on the Hilbert space of the
system. This connection was used to study the enumeration of closed walks of given length and (algebraic) area
on the square lattice. Such walks are generated by the Hofstadter Hamiltonian, with the magnetic field playing the
role of the variable dual to the area, and their properties can be derived
from the study of the secular determinant of the Hamiltonian. The area enumeration generating function
for walks of given length was derived in \cite{nous} in terms of a set of factors
extracted from the secular determinant (the so-called Kreft coefficients \cite{Kreft}), leading to explicit albeit complicated expressions.

An interesting connection was made in \cite{emeis} between a general class of two-dimensional walks and quantum
mechanical particles obeying generalized {\it exclusion statistics} with exclusion parameter $g$ depending on the type of walks ($g\bb=\bb 0$ for bosons, $g\bb=\bb 1$ for fermions, and higher $g$ means a stronger exclusion beyond Fermi).
The relevance of generalized quantum statistics to Calogero particles with inverse-square potential interactions
was first pointed out in \cite{NRBos}. Exclusion statistics was proposed by Haldane \cite{Haldane} as a distillation of the statistical mechanical properties of Calogero-like spin systems. Exclusion statistics also emerges in the context of anyons projected on the lowest Landau level of a strong magnetic field \cite{Das}, and has been extended to more general systems \cite{Poly}. (For a review of exclusion statistics see \cite{gofel}.) Remarkably, the algebraic area considerations of a class
of lattice walks directly map to the statistical mechanics of particular many-body systems with exclusion statistics \cite{emeis}. 

In recent work \cite{DyckPO} the Hamiltonian description of random walks and the exclusion statistics
connection were used to study the generating function of a family of walks referred to as Dyck paths and
their height-restricted generalizations 
$\cite{Stan}{-}\cite{KP}$\footnote{The literature on Dyck and related Motzkin and Lukasiewicz paths is quite extensive. We refer the reader to T. Prellberg's site http://www.maths.qmul.ac.uk/{\tiny$\sim$}tp/
and to the references in \cite{OstJeu} for a comprehensive list of relevant papers.}.
These are walks on a two-dimensional lattice that propagate one step in
the horizontal direction (``time'') and one step either up or down in the vertical direction (``height'') but without dipping below
a ``floor'' at height zero nor exceeding a ``ceiling'' of maximal height. Paths that start and end at the floor are usually termed
``excursions'', while more general paths are ``meanders''.
The Hamiltonian method for forward-moving paths is equivalent to the transition matrix formulation, which has been used in
previous work to calculate the length generating function for such walks. In \cite{DyckPO} these results were extended
to length {\it and} area generating functions for meanders with arbitrary starting and ending points.
Further, using the connection to exclusion statistics, the generating functions were expressed in terms of
statistical mechanical properties of relatively simple particle systems with an equidistant energy spectrum
that are amenable to a full solution by the technique of bosonization. Using a cluster expansion, an alternative form
for the logarithm of the generating functions
was derived in terms of sums over compositions (i.e., ordered partitions) of the integer length of the path that made their polynomial structure explicit.

Motzkin paths are walks that can also propagate by a single step horizontally, in addition to up or down,
with all remaining properties and definitions as in Dyck paths. Their combinatorial enumeration is given by the so-called
Motzkin number \cite{Motzkin} and they have been extensively studied from several points of view (see, e.g., \cite{Flaj}-\cite{JRW}). Various random physical systems can be mapped to
Motzkin-like paths, such as solid-on-solid interfaces, vesicles and, most straightforwardly, polymers.
The floor represents a physical boundary that the polymer cannot cross, while the ceiling confines the polymer
on a strip. The statistical mechanics of these polymers (or other systems) is determined by the combinatorics of Motzkin
paths and, in general, exhibits phase transitions between diffuse and localized states.


Several combinatorial properties and generating functions of Motzking paths have been considered and studied.
The full length {\it and} area generating function for Motzkin meanders, however, has apparently not been calculated
(\cite{RSOS} comes closest to that goal, evaluating generating functions for excursions, that is, Motzkin paths
starting and ending on the floor). In this paper we apply the techniques of \cite{DyckPO} to calculate the length and area
generating function of Motzkin paths,
with different weights assigned to each kind of step (up, horizontal or down) and a set of additional variables probing their
boundary properties.
Although the basic methodology is the same as for Dyck paths, the application of exclusion statistics techniques in the
Motzkin case presents additional challenges that require some new tricks. Nevertheless, the full generating function is derived
in terms of determinants, related to exclusion-2 statistical systems and calculated via bosonization.

In the next section we set up the Hamiltonian description of Motzkin paths and express their
generating functions in terms of matrix elements of the propagator, while in section 3 we derive the basic determinant
formula for the generating functions, including additional variables (markers) monitoring their passage and time spent
on the floor or ceiling, and examine several special cases.
In section 4 we introduce the anisotropic two-step Dyck process that generates
Motzkin paths, review the exclusion statistics connection and use it to express the basic building block of the generating
functions, i.e., the secular determinant of the two-step process, in terms of grand partition functions
and Chebyshev polynomials. In section 5 we use
cluster decomposition techniques to derive expressions for the logarithm of the generating functions of Motzkin paths.
We conclude in section 6 with some remarks on previous work and directions for future research.

This is an opportune moment to log an apology to any mathematician readers. The introduction of concepts such as
quantum exclusion statistics and bosonization may present for them an additional burden of familiarization and supplant a purely
mathematical treatment that would eschew such schemes and jargon. We feel, nevertheless, that this approach, apart from reflecting the parochial
point of view of the author, may add some physical context to the calculations and could be a source of inspiration
and insight to those approaching the problem from other vantage points. (This paper is also written in a
narrative style, rather than the {\it Proposition/Theorem} format canonical to mathematics publications, for which
we make no apologies.)

\section{Motzkin path Hamiltonian}

Motzkin paths are forward-moving random walks on a square lattice on the first quadrangle of the plane
consisting of points $(i,j)$, $i = 0,1,2,\dots$, $j = 0,1,2,\dots k$. The walk in each step moves one horizontal unit, $i\to i+1$,
and either 0 or 1 vertical units in either direction, $j \to j, j\pm 1$. (Walks where the horizontal step $j \to j$ is forbidden
are called Dyck paths.) Paths can never dip below a lowest level (``floor'') $j=0$ nor exceed a maximum height
(``ceiling'') at $j=k$.
Relevant quantitative features of the path are its starting and finishing heights $m$ and $n$,
respectively, the number of upward, horizontal and downward steps $l_u,l_h,l_d$, as well as the total area vertically under the walk $A$, the total length of the walk being $l=l_u+l_h+l_d$ (see fig.\ref{Motzkin path}).

An object of special
interest is the generating function of walks ``packaging'' the above quantitative features, defined as
\be
G_{k,mn} (z_u,z_h,z_d,q) = \sum_{l_u,l_h,l_d,A}^\infty z_u^{l_u} z_h^{l_h} z_d^{l_d} q^A N_{k,mn;l_u,l_h,l_d,A}
\ee
with $N_{k,mn;l_u,l_h,l_d,A}$ the number of walks with the given parameters. We can eliminate one of the dual variables
$z_u , z_d$, right away: clearly $l_u - l_d = n-m$, and
\be
G_{k,mn} (z_u,z_h,z_d,q) = z_d^{m-n} \sum_{l_u,l_h,A}^\infty (z_u z_d)^{l_u} z_h^{l_h} q^A N_{k,mn;l_u,l_h,l_u+m-n,A}
\ee
So the dependence on $z_d$ is trivial, the relevant variable being $z_u z_d$. The choice $z_d=1$ could have been made,
but we prefer the choice $z_u = z_d = z$, making the Hamiltonian symmetric and the generating function depend
only on $z,z_h,q$. The calculation of $G_{k,mn} (z,z_h,q)$ will be the main focus of this paper.

Discrete forward-moving paths can be described in terms of a Hamiltonian (transition matrix) acting on a Hilbert space
of dimensionality equal to the number of states (vertical positions) that the path can visit. Its structure encodes the
allowed steps and keeps an account of the quantitative properties of the paths.
The Hamiltonian for Motzkin paths of maximum height $k$ can be expressed as the $(k\bb +\bb 1)$-dimensional matrix
\be
{H}_k = 
\begin{pmatrix}
z_h\;\;& z q^{1/2}\;\, & 0\;\; & 0\;\; \cdots\;\; & 0\;\; & 0\; \\
z q^{1/2} \;\;& z_h q \;\;&z q^{3/2}\;\;& 0\;\; \cdots \;\;& 0 \;\;& 0\; \\
0 \;\;& z q^{3/2}\;\; & z_h q^2\;\; &z q^{5/2} \;\cdots\;\; & 0 \;\;& 0\; \\
\vdots \;\;& \vdots \;\;& \vdots \;\;& \;\ddots & \vdots\;\; & \vdots\; \\
0\;\; & 0 \;\;& 0 \;\; &0\;\;  \cdots& z_h q^{k-1}\;\; & z q^{k-1/2} \;\\
0\;\;& 0 \;\;& 0 \;\; &0 \;\;\cdots & z q^{k-1/2}\;\;& z_h q^{k}\; \\
\end{pmatrix}
\label{hamk}\ee
$H_k$  includes the parameters $z,z_h$ and $q$ of the generating function and is symmetric (and
Hermitian, for real parameters, although this will be of no import for our considerations).
In previous work on Dyck paths \cite{DyckPO} the variable $z$ dual to the length was an external multiplicative
parameter, but here we prefer to include all dual variables in the Hamiltonian.

We will assume that the Hamiltonian acts on a Hilbert space with basis elements $\ket{j}$ and produces a single step.
Repeated action of $H_k$ produces a superposition of all possible walks with weights as they appear in the generating
function. Specifically, the left-action of $H_k$ on the dual state $\bra{j}$ produces the superposition
\be
\bra{j} H_k = z q^{j+1/2} \bra{j+1} + z_h q^j \bra{j} + z q^{j-1/2} \bra{j-1}
\label{weights}\ee
with $\ket{k+1} \equiv 0 \equiv \ket{-1}$.
(We chose $H_k$ to act on the left to match time evolution from left to right to the ordering of operators.)
Mapping the vertical position $j$ to the Hilbert space element $\ket{j}$, we can interpret the action of $H_k$
as producing a unit vertical step either up to $\ket{j+1}$, or horizontally to $\ket{j}$, or down to $\ket{j-1}$.
A single application of $H_k$
corresponds to a unit step $i \to i+1$. The vertical area under such a step $(i,j) \to (i+1,j+\Delta j)$
measured in units of lattice plaquettes is $a= \bigl[ j+ (j+\Delta j)\bigr]/2 = j+\half,j,j-\half$ for
an up, horizontal or down step, respectively, and therefore the weighting factors in (\ref{weights}) are $\za q^{a}$
with $z_i=z,z_h$ depending on the type of step.
The repeated application $H_k^l$, then, produces a superposition of all possible Motzkin paths of $l$ steps
starting at height $j$, each path weighted by a factor arising from the products of the above coefficients in each step;
that is, by a factor $z^{l_u} z_h^{l_h} z^{l_d} q^{A}$ with $A$ the total area under the path.
States $\ket{0}$ and $\ket{k}$ (corresponding to lattice points $(i,0)$ and $(i,k)$) constitute a `floor' and  a `ceiling'.

{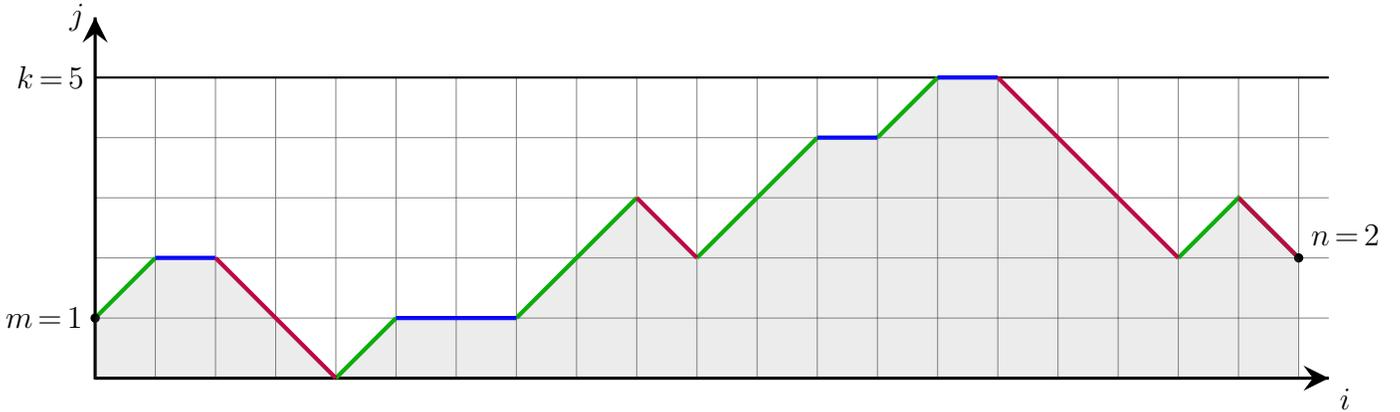
\begin{figure}
{\hskip -1.2cm
\begin{tikzpicture}[scale=0.8]

\draw[help lines, gray] (0,-0.02) grid (20.5,5.01);

\tikzset{big arrow/.style={decoration={markings,mark=at position 1 with {\arrow[scale=3,#1,>=stealth]{>}}},postaction={decorate},},big arrow/.default=black}

\draw[very thick,-] (0,0) -- (20.5,0);
\draw[very thick,-] (0,-0.02) -- (0,6);
\draw[thick,-] (0,5) node[left] {$k\bb=\bb 5$} -- (20.5,5);


\draw[big arrow] (0,0) -- (20.5,0) node[below right] {$i$};

\draw[big arrow] (0,-0.02) -- (0,6) node[left] {$j$};

\draw[ultra thick,black!30!green,-](0,1)node[left] {{\color{black} $m\bb=\bb 1$}}--(1,2);
\draw[ultra thick,black!30!green,-](4,0)--(5,1); \draw[ultra thick,black!30!green,-] (7,1)--(9,3); \draw[ultra thick,black!30!green,-](10,2)--(12,4); \draw[ultra thick,black!30!green,-](13,4)--(14,5);
\draw[ultra thick,black!30!green,-](18,2)--(19,3)--(20,2);
\draw[ultra thick,blue,-](1,2)--(2,2); 
\draw[ultra thick,blue,-](5,1)--(7,1); 
\draw[ultra thick,blue,-](12,4)--(13,4);
\draw[ultra thick,blue,-](14,5)--(15,5);
\draw[ultra thick,purple,-](2,2)--(4,0); \draw[ultra thick,purple,-](9,3)--(10,2); \draw[ultra thick,purple,-](15,5)--(18,2);
\draw[ultra thick,purple,-](19,3)--(20,2)node[above right] {\color{black} $n\bb=\bb 2$};

\fill (0,1) circle (2.2pt);
\fill (20,2) circle (2.2pt);

\draw[transparent,ultra thin,fill=gray,fill opacity=0.15](0,0)--(0,1)--(1,2)--(2,2)--(4,0)--(5,1)--(7,1)--(9,3)--(10,2)--(12,4)--(13,4)--(14,5)--(15,5)--(18,2)--(19,3)--(20,2)--(20,0)--(0,0);

\end{tikzpicture}
}
\vskip -1cm
\caption{\small{A typical Motzkin path (meander) for $k=5$, starting at $m=1$ and ending at $n=2$, with
$l_u = 8$ (green) up-steps, $l_h = 5$ (blue) horizontal steps and $l_d = 7$ (red) down-steps for a total length of $l=20$ steps. The area under it is 49.5 plaquettes (shaded gray).}}\label{Motzkin path}
\end{figure}}


The above correspondence of Motzkin paths with the action
of $H_k^l$ makes it clear that the $m,n$ matrix element of $H_k^l$ reproduces the sum of walks with $l$ steps starting at
height $m$ and ending at height $n$ weighted by their area and number of each type of steps
\be
\bra{m} H_k^l \ket{n} = \sum_{l_u,l_h,l_d,A}^\infty z^{l_u+l_d} z_h^{l_h} q^A N_{k,mn;l_u,l_h,l_d,A} \,\delta(l_u+l_h+l_d -l)
\ee
and the full generating function becomes a matrix element of the ``propagator'' $(1-H_k)^{-1}$
\be
G_{k,mn} (z,z_h,q) = \sum_{l=0}^\infty \bra{m} H_k^l \ket{n} = \bra{m} (1- H_k )^{-1} \ket{n}
\ee
(we assumed small enough $|\za|$ and $|q|$ for convergence of the sums). It is clear from this form that the generating function satisfies the convolution property
\be
G_{k,mn} (z,z_h,q)  = \sum_{j=0}^k G_{k,mj} (z,z_h,q) G_{k,jn} (z,z_h,q) 
\ee
For later convenience, we will adopt the simplifying (and hopefully intuitive) convention that indices $k\bb =\bb\infty$
(no ceiling) and $mn\bb =\bb 00$ (excursions) are omitted, while indices $mn=kk$ (paths `hanging' from the ceiling)
are replaced by overbar; that is,
\be
G_{\infty,mn}  = G_{mn} ~ ,~~ G_{k,00}  = G_k ~,~~ G_{k,kk} = {\overline G}_k ~,~~
G_{\infty,00}  = G
\label{conventions}\ee

In the sequel we will show that the above generating function can be expressed as a rational expression of determinants,
and will evaluate these determinants by connecting them to generalized quantum exclusion statistics of order 2 and using
bosonization.

\section{Determinant formula for the generating function}\label{detsec}

Our goal is the evaluation of the matrix elements of the propagator matrix $(1- H_k )^{-1}$ that appear
in the generating function.

\subsection{Basic result}

The derivation proceeds much along the lines of the corresponding calculation for Dyck paths \cite{DyckPO}.
We define the secular matrix
\be
\hskip -0.1cm D_k (\za,q) = 1\bb-\bb H_k
=\begin{pmatrix}
1\bb\bb-\bb\bb z_h\;\;& -z q^{1/2} \;\;& 0 \;& 0 ~\cdots & \bb\bb 0 \;& 0 \\
-z q^{1/2} \;\;&1\bb\bb-\bb\bb z_h q \;&-z q^{3/2}\;& 0 ~\cdots &\bb\bb 0 \;& 0 \\
0 \;& -z q^{3/2} \;&1\bb\bb-\bb\bb z_h q^2 \;&-z q^{5/2} \cdots & \bb\bb 0 \;& 0 \\
\vdots & \vdots & \vdots & \ddots &\bb\bb \vdots \;& \vdots \\
0 & 0 & 0  &0 ~\cdots&\bb\bb 1\bb\bb-\bb\bb z_h q^{k-1} \;& -z q^{k-1/2} \\
0& 0 & 0  &0~ \cdots &\bb\bb -z q^{k-1/2}\;&1\bb\bb-\bb\bb z_h q^{k} \\
\end{pmatrix}
\label{Dk}\ee
with $\za$ denoting collectively $z,z_h$, as well as its determinant and matrix elements of its inverse (generating function)
\be
F_k (\za,q) = \det D_k (\za,q) ~,~~~
G_{k,mn} (\za,q) = \bra{m} D_k (\za,q)^{-1} \ket{n}
\ee
Clearly $F_0 (\za,q)=1\bb-\bb z_h$, and we also define  $F_{-1} (\za,q) \bb=\bb 1$ and $F_k (\za,q) \bb=\bb 0$ for $k\le -2$.

$G_{k,mn}$ is calculated by the standard formula for the elements of the inverse of a matrix in terms of its
cofactors. Applied to matrix $D_k (\za,q)$ it yields
\be
\bra{m} D_k (\za,q)^{-1} \ket{n}  = (-1)^{m-n} \; {\det D_k (\za,q)_{(nm)} \over \det D_k (\za,q)}
\ee
where the complement $D_k (\za,q)_{(nm)}$ is the matrix $D_k (\za,q)$ with the $n^{\text {th}}$ row and
$m^{\text {th}}$ column removed.

The denominator in the right-hand side is $F_k (\za,q)$. The remaining determinant of $ D_k (\za,q)_{(nm)}$ can be
related to simple secular determinants. First, observe that the secular matrix with its first $n$ rows
and columns truncated, denoted $D_k (\za,q)_{[n]}$, is related to the secular matrix for a reduced $k$. Specifically,
\be
D_k (\za,q)_{[n]} \bb = \bb \begin{pmatrix}
1\bb-\bb z_h q^n& -z q^{n+1/2} & 0\;\; ~~~\cdots & \bb\bb\bb 0 & \bb\bb 0 \\
-z q^{n+1/2} & 1\bb-\bb z_h q^{n+1} &\bb -z q^{n+3/2}  \cdots & \bb\bb\bb 0 & \bb\bb 0 \\
0 & -z q^{{n+3/2}} & 1\bb-\bb q^{n+2} ~~~ \cdots & \bb\bb\bb 0 & \bb\bb 0 \\
\vdots & \vdots  & \;~~ \; \ddots & \bb\bb\vdots &\bb \bb\bb \vdots \\
0 & 0 &  0 \; ~~~\cdots & \bb\bb\bb 1\bb-\bb z_h q^{k-1} &\; - z q^{k-1/2} \\
0& 0  &0 \;~~~\cdots &\bb\bb\bb -z q^{k-1/2}&\bb 1\bb-\bb z_h q^k \\
\end{pmatrix}
\bb= D_{k-n} (\za q^n , q)
\label{trunc}
\ee
Assuming, for now, $m\le n$, it is easy to see that the complement $D_k (\za,q)_{(nm)}$ becomes
block-triangular of the form
\be
D_k (\za,q)_{(nm)} = 
\begin{pmatrix}
D_{m-1} (\za,q) &\; 0 & 0 & \\
  &  &  \\
A &\; Q &0 \\
  &  &  \\
B & \; C &  \hskip 0.2cm D_k (\za,q)_{[n+1]}\bb\bb \\
\end{pmatrix}
\label{DQD}\ee
with $Q$ a lower-diagonal matrix. The crucial property of $H_k$ and $D_k (\za ,q)$ that leads to this
form of $D_k (\za ,q)_{(nm)}$ and $Q$ is the fact that they are paradiagonal with two off-diagonals flanking the
diagonal, a feature shared with Dyck paths. This block diagonal form of $D_k (\za ,q)_{(nm)}$ implies
\be
\det D_k (\za,q)_{(nm)} = \det \bb D_{m-1} (\za,q)\, \det\bb Q \, \det\bb D_k (\za,q)_{[n+1]}
\label{detQ}\ee
$Q$ is lower-triangular with diagonal elements $-z q^{m+1/2} , -z q^{m+3/2} , \dots , -z q^{n - 1/2}$
for $m<n$, and is completely absent if $m=n$.
Therefore,
\be
\det Q = (-z)^{n-m} \, q^{n^2 - m^2 \over 2}
\ee
Putting everything together, and using (\ref{trunc}), we finally obtain
\bea
G_{k,mn} (\za,q) &=& z^{n-m}\,  q^{n^2 - m^2 \over 2} \, {F_{m-1} (\za,q)\, F_{k-n-1} (\za q^{n+1} ,q) \over F_k (\za,q)}
~,~~ n\ge m \nonumber \\
&=& z^{m-n}\,  q^{m^2 - n^2 \over 2} \, {F_{n-1} (\za,q)\, F_{k-m-1} (\za q^{m+1} ,q) \over F_k (\za,q)}
~,~~ n \le m
\label{typara}\eea
the second formula following from the symmetry $G_{k,mn} (\za,q) = G_{k,nm} (\za,q)$. The formula holds for all
values of $k,m,n$ under the conventions for $F_k$ for negative values of $k$.

Formula (\ref{typara}) is our first main result. It is practically identical in form to the corresponding result for Dyck paths,
the difference between the two kinds of paths being in the properties of the secular determinant $F_k (\za,q)$.
In the next section we will calculate this determinant by relating it to a two-step alternating Dyck process, which will allow us
again to express it as the grand partition function of a quantum exclusion statistics system but for a more involved spectrum
than the one in the Dyck case.

\subsection{Special cases}

A few interesting special cases are worth recording. For diagonal height-restricted paths starting and ending at
the same height $m=n$ (higher excursions) we have
\be
G_{k,nn} (\za,q) = {F_{n-1} (\za,q) F_{k-n-1} (\za q^{n+1} ,q) \over F_k (\za,q)}
\ee
For excursions, in particular, $G_{k,00} \bb =\bb G_k$ becomes the ratio of determinants
\be
G_k (\za,q)= {F_{k-1} (\za q ,q) \over F_k (\za,q)}
\label{Gexcur}\ee
and for `dual' paths hanging from the ceiling,
$G_{k,kk} (\za,q) \bb=\bb {\overline G}_k (\za,q)$ becomes the simpler expression
\be
{\overline G}_{k} (\za,q) = {F_{k-1} (\za ,q) \over F_k (\za,q)}
\ee
Finally, for unrestricted Motzkin excursions (higher and floor ones) we have
\be
G_{nn} (\za,q) = F_{n-1} (\za,q) G (\za,q)~,~~
G (\za,q) = {F (\za q ,q) \over F (\za,q)}
\ee

\subsection{Duality and recursion relations}\label{recursions}

The secular matrix and determinant satisfy the duality relation
\be
D_k (\za q^k, q^{-1}) = \sigma D_k (\za, q)\, \sigma ~~~ \Rightarrow ~~~ 
F_k (\za q^{k} , q^{-1} ) = F_k (\za,q)
\label{duality}\ee
where $\sigma_{mn} = \delta_{m+n,k}$ is the reflection matrix. This expresses the symmetry of walks under vertical reflection around the median line at $k/2$, mapping $\ket{n} \to \ket{k-n}$.
(\ref{duality}) implies the corresponding duality relation for generating functions
\be
G_{k,mn} (\za,q) = G_{k;k-n,k-m} (\za q^{k}, q^{-1} )
\ee

Several generating function recursion relations can be deduced directly from the form itself of (\ref{typara}), irrespective of
the form of $F_k (\za,q)$. For instance,
\bea
G_{k,mn} (\za,q)  &=& z q^{n-1/2}\, G_{k;m,n-1} (\za,q) \, G_{k-n} (\za q^n ,q) ~~~~~~~~~~~~(m<n) \nonumber \\
&=&\, z q^{l+1/2}\; G_{k;l+1,n} (\za,q) \, {G}_{l;m,l} (\za,q)~\,~~~~~~~~~~~~(m \le l < n) \label{firstpass}
\eea
Further recursion relations, more specific to Motzkin paths, can be derived by expanding $\det D_k (\za,q)$ in terms of its top row, as in the Dyck path case. We obtain
\be
F_k (\za,q) = (1-z_h ) F_{k-1} (\za q ,q ) - z^2 q F_{k-2} (\za q^2 , q)
\label{ex}\ee
which leads to corresponding relations for $G_{k,nm} (\za,q)$. Several such relations can be written, and we choose
to present two: for generic paths, applying (\ref{ex}) to the term $F_{k-n-1} (\za q^{n+1} ,q)$ in (\ref{typara}) yields
\be
(1-z_h q^n ) G_{k,mn} (\za,q) = z q^{n-1/2}\, G_{k;m,n-1} (\za,q) + z q^{n+1/2} \,  G_{k;m,n+1} (\za,q) ~~~~~(m<n<k)
\ee
and for excursions, dividing (\ref{ex}) by $F_k (\za,q)$ yields
\be
(1- z_h ) G_k (\za , q) = 1 + z^2 q \, G_{k-1} (\za q,q) \, G_k (\za,q) 
\label{concur}\ee
All the above recursion relations admit geometric interpretations in terms of decomposing
paths into their parts. Figure \ref{Geometric} demonstrates the geometric significance of (\ref{concur}),
which generalizes a similar construction for Dyck paths.

{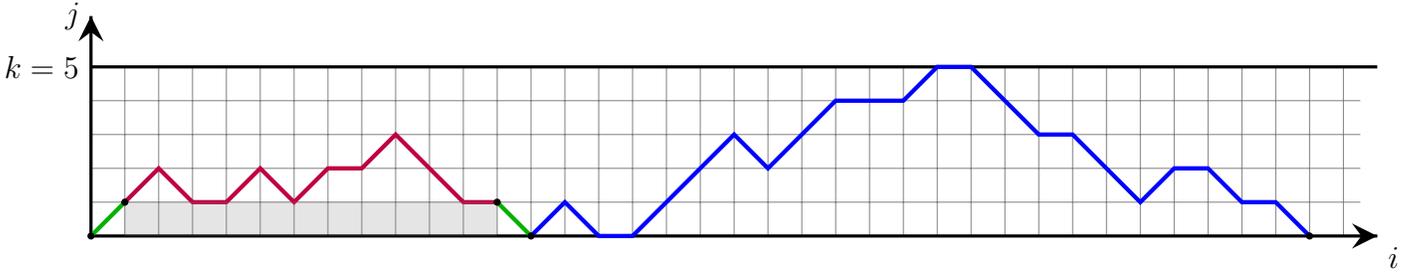
\begin{figure} \vskip -1.8cm
{\hskip -1.5cm
\begin{tikzpicture}[scale=0.45]
\draw[help lines, gray] (0,-0.02) grid (37.5,5.01);

\tikzset{big arrow/.style={decoration={markings,mark=at position 1 with {\arrow[scale=3,#1,>=stealth]{>}}},postaction={decorate},},big arrow/.default=black}

\draw[very thick,-] (0,0) -- (38,0);
\draw[very thick,-] (0,-0.02) -- (0,6.5);
\draw[very thick,-] (0,5) node[left] {$k=5$} -- (38,5);

\draw[big arrow] (0,0) -- (38,0) node[below right] {$i$};
\draw[big arrow] (0,-0.02) -- (0,6.5) node[left] {$j$};

\draw[ultra thick,purple,-](1,1)--(2,2)--(3,1)--(4,1)--(5,2)--(6,1)--(7,2)--(8,2)--(9,3)--(11,1)--(12,1);
\draw[ultra thick,black!30!green,-](0,0)--(1,1); \draw[ultra thick,black!30!green,-](12,1)--(13,0);
\draw[ultra thick,blue,-](13,0)--(14,1)--(15,0)--(16,0)--(19,3)--(20,2)--(22,4)--(23,4)--(24,4)--(25,5)--(26,5)--(28,3)--(29,3)--(31,1)--(32,2)--(33,2)--(34,1)--(35,1)--(36,0);
\draw[transparent,ultra thin,fill=gray,fill opacity=0.2](1,0)--(1,1)--(12,1)--(12,0)--(1,0);

\fill (0,0) circle (3pt);\fill (1,1) circle (3pt);\fill (12,1) circle (3pt);
\fill (13,0) circle (3pt);\fill (36,0) circle (3pt);

\end{tikzpicture}
}\vskip -0.3cm
\caption{\small{An interpretation of (\ref{concur}), written as 
$G_k (\za ) = z_h G_k (\za) +{\color{black!40!green} {z^2 q}} \, {\color{purple}{G_{k-1} (\za q)}} \, {\color{blue}{G_k (\za)}} +1$,
 as a ``first passage'' equation. The first step
of excursions starting at $(0,0)$ can be either horizontal or up. If it is horizontal (not shown in the figure), it contributes a
factor of $z_h$ and the remaining path is also a general excursion, accounting for the term $z_h G_k (\za)$.
If the first step is up (such a path of length 26 is depicted above), it can
be decomposed into a path returning to the floor $j=0$ {\it for the first time} [first part (green \& red)
of path] and the remaining (blue) arbitrary excursion. For paths of length at least two, the first-passage path has one first
step and one last step (green), each contributing a factor of $z q^{1/2}$.
The remaining upper (red) part never dips below $j\bb =\bb 1$ and can be interpreted as an excursion, but with a
ceiling reduced by 1 and an area increased by its length (shaded plaquettes), contributing the factor $G_{k-1}
(\za q)$, the shift $\za \to \za q$ accounting for the extra area. The remaining (blue) path contributes $G_k (\za)$.
Finally, the trivial path of length zero cannot be decomposed and contributes the term 1.
Relations (\ref{firstpass}) also admit a similar first-passage interpretation.}}\label{Geometric}
\end{figure}}

\subsection{Top and bottom event markers}\label{tobo}

Before proceeding to the calculation of the secular determinant in the next section, we derive the expression of
a generalization of the generating function that also keeps track of the times a path ``hits''
the floor $n=0$ (a `touch-down'), the total time (in step units) that it spends on the floor (`creep-down') the number
of times it hits the ceiling $n=k$ (`touch-up') and the total time it spends on the ceiling (`creep-up'). E.g.,
the path of fig.\ref{Geometric} has three touch-downs, 1 creep-down, 1 touch-up and 1 creep-up. These are examples
of various `markers' that we can add to monitor local properties of the paths.

Weighing each touch-down with a factor of $t$, each creep-down with a factor of $s$, each touch-up with a factor
of $\T$ and each creep-up with a factor of $\s$, the generating function for paths from $m$ to $n$ with length $l$,
area $A$, $l_h$ horizontal steps, $a$ touch-downs, $b$ creep-downs, $c$ touch-ups and $d$ creep-ups becomes
\be
\tG_{k,mn} (t,s;\T,\s\,|\za,q) = \sum_{A,l,a,b,c,d=0}^\infty t^a\, s^b\, \T^c\, \s^d \, z^{l-l_h} z_h^{l_h} \, q^A 
N_{k,mn;l,l_h,A,a,b,c,d}
\ee
(we continue setting $z_u = z_d = z$).
Clearly $\tG_{k,mn} (1,1;1,1|z_i,q) = G_{k,mn} (\za,q)$.
 
This generalization can be  implemented in our Hamiltonian framework with minor modifications.
Consider the Hamiltonian $\tH_k$ ($k \ge 1$)
\be
{\tH}_k = 
\begin{pmatrix}
s z_h\;\;& z q^{1/2}\;\, & 0\;\; & 0\;\; \cdots\;\; & 0\;\; & 0\; \\
t z q^{1/2} \;\;& z_h q \;\;&z q^{3/2}\;\;& 0\;\; \cdots \;\;& 0 \;\;& 0\; \\
0 \;\;& z q^{3/2}\;\; & z_h q^2\;\; &z q^{5/2} \;\cdots\;\; & 0 \;\;& 0\; \\
\vdots \;\;& \vdots \;\;& \vdots \;\;& \;\ddots & \vdots\;\; & \vdots\; \\
0\;\; & 0 \;\;& 0 \;\; &0\;\;  \cdots& z_h q^{k-1}\;\; & \T z q^{k-1/2} \;\\
0\;\;& 0 \;\;& 0 \;\; &0 \;\;\cdots & z q^{k-1/2}\;\;& \s z_h q^{k}\; \\
\end{pmatrix}
\label{hamtk}\ee
$\tH_k$ is the same as $H_k$ but with the $\ket{0}\bb\bra{0}$ element multiplied by $s$, the $\ket{1}\bb\bra{0}$
element multiplied by $t$, the $\ket{k-1}\bb\bra{k}$ element multiplied by $\T$, and the 
$\ket{k}\bb\bra{k}$ element multiplied by $\s$.
It should be obvious that $\tH_k^l$ counts area-weighted paths of length $l$, as before, but also multiplies by an
appropriate factor each up- or down- event. Therefore, as before,
\be
\tG_{k,mn} (t,s,\T,\s\,|\za,q) = \sum_{l=0}^\infty \bra{m} \tH_k^l \ket{n} = \bra{m} (1- \tH_k )^{-1} \ket{n}
\ee
Note that $\tH_k$ is not symmetric for $t,\T\neq 1$.
The asymmetry is due to the fact that paths entering or exiting the floor or the ceiling are weighted differently,
and implies $\tG_{k,mn} \neq \tG_{k,nm}$ if $m,n=0,k$. Nevertheless, we can render $\tG_{mn}$ fully symmetric
by assigning an extra weight $t$ to paths starting from the floor
and an extra weight $\T$ to paths starting from the ceiling, and we shall adopt this convention. (Another alternative,
often used in the literature, would be to not count the final touch-down or touch-up of paths ending at $0$ or $k$.
However, as we shall see, our symmetrization convention leads to more compact expressions.)

Denoting
\be
\tF_k (t,s;\T,\s\,|\za,q) = \det (1-\tH_k )
\ee
a calculation entirely analogous to the $t=s=\T=\s=1$ case yields for $2\bb\le\bb m\bb \le\bb n\bb\le\bb k-2$
\be
\tG_{k,mn} (t, s;\T,\s\,|\za,\bb q) =
z^{n-m}\,  q^{n^2 - m^2 \over 2} \, {\tF_{m-1} (t,s;1,1|\za,q) \, 
\tF_{k-n-1} (1,1;\T,\s\,|\za q^{n+1} ,q) \over \tF_k (t,s;\T,\s\,|\za,q)}
\label{Typara}\ee
$\tH_k$ and $\tF_k$ are not defined for $k<1$. Nevertheless, if we define
\be
\tF_0 = t\bb+\bb\T\bb -\bb t \s z_h\bb\bb -\bb\T s z_h\bb\bb -\bb t\T\bb +\bb t\T z_h~
,~~~\tF_{-1} \bb=\bb t \T~ ,~~~
\tF_k =0 ~ (k\le \bb -2)
\label{convy}\ee
then (\ref{Typara}) becomes valid for all values of $n,m$: for $m=1$ or $n=k-1$, $\tF_0$ as defined in (\ref{convy})
reproduces the correct factors $1-s z_h$ or $1-\s z_h q^k$ arising from the evaluation of the corresponding matrix elements
of $(1-\tH_k )^{-1}$. For $n=k$ the last factor in the numerator is absent, but the matrix element
includes an extra factor of $\T$ arising from the structure of the $Q$ matrix
appearing in the analogs of (\ref{DQD},\ref{detQ}), which involves the $\bra{k-1} \tH_k \ket{k}$ element of $H_k$,
and this factor is reproduced by $\tF_{-1} (1,1;\T,\s)$ in (\ref{Typara}). Finally, for $m=0$ the first factor
in the numerator is absent, but by our symmetrization convention we must include an extra factor of $t$,
which is reproduced by $\tF_{-1} (s,t;1,1)$. Therefore, conventions (\ref{convy}) make formula
(\ref{Typara}) valid for the full range of values of $m,n$.
 
To relate $\tF_k$ to $F_k$, we expand the determinant in terms of its top row. We obtain, for $k\ge 1$,
\be
\tF_k (t,s;\T,\s\,|\za,q) = (1-s z_h) \tF_{k-1} (1,1;\T,\s\,|q\za,q)  - t\, z^2 q \tF_{k-2}  (1,1;\T,\s\, |q^2 \za,q) 
\label{tex}\ee
and combining with the same formula for $s=t=1$ yields
\be
\tF_k (t,s;\T,\s\,|\za,q) = t\, \tF_k  (1,1;\T,\s\,|\za,q)  + (1\bb-\bb t\bb-\bb s z_h\bb +\bb t z_h) \tF_{k-1} (1,1;\T,\s\,|q\za,q) 
\label{trec}\ee
Similarly, expanding $\tF_k$ in terms of its bottom row gives
\be
\tF_k (t,s;\T,\s;\,|\za,q) = (1-\s z_h q^k) \tF_{k-1} (t,s;1,1|\za,q)  - \T\, z^2 q^{2k-1} \tF_{k-2}  (t,s;1,1|\za,q) 
\label{Tex}\ee
and combining with the same formula for $\T=\s=1$ yields
\be
\tF_k (t,s;\T,\s\,|\za,q) = \T\, \tF_k  (t,s;1,1|\za,q)  + (1\bb-\bb\T\bb-\bb\s z_h  q^k\bb+\bb \T z_h q^k) \tF_{k-1} (t,s;1,1|\za,q) 
\label{Trec}\ee
Finally, applying formula (\ref{trec}) for $\T\bb =\bb\s \bb=\bb 1$ and inserting in (\ref{Trec}) (or vice versa) gives
\bea
&&\tF_k (t,s;\T,\s\,|\za,q) = t\, \T\, F_k (\za,q) + t (1\bb-\bb\T\bb-\bb\s z_h q^k\bb +\bb \T z_h q^k ) F_{k-1} (\za,q) \label{tT}\\
&& + \T (1\bb-\bb t\bb -\bb s z_h \bb+\bb t z_h ) F_{k-1} (\za q, q) + (1\bb-\bb t\bb-\bb s z_h\bb +\bb t z_h ) 
(1\bb-\bb\T\bb-\bb\s z_h q^k \bb+\bb \T z_h q^k ) F_{k-2} (\za q ,q)
\nonumber\eea
Note that, with the definitions (\ref{convy}), the above formula holds for all $k$, including $k<1$.
(This motivates the somewhat unintuitive form of $\tF_{-1}$.)

We have thus related the secular determinant of the process with markers to that of the unmarked
process, which will be calculated in the next section. Applying the above formula for the terms in (\ref{Typara}),
and also using the expression (\ref{typara}) for $G_{k,mn}$, leads to the relation for $\tG_{k,mn} = \tG_{k,nm}$
for $m\le n$
 \be
\tG_{k,mn} (t,s;\T,\s) = {\bigl[ t\bb+\bb A_0 (t,\bb s)\, G_{m-1} \bigr]
\bigl[\T\, G_{k,mn} \bb+\bb 
{A}_k (\T,\bb\s)\,\, {\overline G}_{k} \, G_{k-1,mn} \bigr]
\over {A}_k(\T,\bb\s) \bigl[t \bb+\bb A_0 (t,\bb s)\, G_{k-1}\bigr]{\overline G}_{k}  + 
\T\bigl[ t\bb+\bb A_0 (t,\bb s) \, G_k \bigr] }
\label{hag}\ee
where, for brevity, we suppressed the (common) arguments $\za,q$ and defined
\be
A_r (t,s) = 1 - t + (t-s)\, z_h \, q^r
\ee

Formula (\ref{hag}) is our second main result and expresses the generating function $\tG_{k,mn}$ in terms of $G_{k,mn}$.
It is conceivable that
a relation between $\tG_{k,mn}$ and $G_{k,mn}$ could be obtained combinatorially, with arguments
similar to the ones of fig.\ref{Geometric}. However, the rather complicated form of (\ref{hag}) suggests that such an argument
would be quite convoluted. Our Hamiltonian approach allowed for a relatively straightforward derivation of this
relation without combinatorial ingenuity.

We conclude with a few remarks and special cases. $\tG_{k,mn}$ satisfies the floor-ceiling duality relation
\be
\tG_{k,mn} (t,s;\T,\s\,|\za,q) = \tG_{k;k-m,k-n} (\T,\s;t,s\,|\za q^k,q^{-1})
\ee
This is a consequence of the duality relation for $\tF_k$
\be
\tF_{k} (t,s;\T,\s\,|\za,q) = \tF_{k} (\T,\s;t,s\,|\za q^k,q^{-1})
\ee
which is obvious from the form of $\tH_k$ (\ref{hamtk}) but also follows from (\ref{tT}) and the corresponding
duality (\ref{duality}) for $F_k$.
Expression (\ref{hag}) does not appear to respect this duality, as it does not look symmetric in $t,s$
and $\T,\s$. However, this is an artifact of the specific form of the expression; using identities as derived in section \ref{recursions},
and the fact that (\ref{hag}) is valid for $m\le n$, duality is restored.

For $z_h \bb=\bb 0$, $\tG_{k,mn}$ has no dependence on
$s$ and $\s$. Indeed, for $z_h =0$ the process degenerates to Dyck paths, which cannot creep over the floor
nor over the ceiling. For $\T \bb=\bb \s \bb=\bb 1$, only touch-downs and creep-downs are monitored.
$\tG_{k,mn}$ becomes
\be
\tG_{k,mn} (t,s;1,1) = G_{k,mn} {t+[1\bb-\bb t+(t\bb-\bb s) z_h ] \,G_{m-1} \over t+[1\bb-\bb t+(t\bb-\bb s) z_h ] \,G_k}
\label{down}\ee
and for $z_h = 0$ the above formula reproduces the result for Dyck paths with touch-downs obtained in \cite{DyckPO}.

For $s=t$, $\s = \T$, only the total number of lattice sites on the floor and on the ceiling are monitored.
Remarkably, (\ref{hag}) does not involve $z_h$ in that case, other than the implicit dependence through $G_{k,mn}$,
so the same formula remains valid for the case of Dyck paths.

For $k=\infty$ (no ceiling), ${\overline G}_k = 0$. The dependence on $\T,\s$ drops, as expected, and
$\tG (t,s)$ is given by (\ref{down}) with $G$ (the unrestricted excursions generating function) instead of $G_k$
in the denominator.

Finally, for
\be
s = t +{1-t \over z_h} ~,~~ \s = \T +{1-\T \over z_h}q^k  ~~~ \Rightarrow ~~~ \tG_{k,mn} = G_{k,mn}
\ee
Remarkably, there is a two-parameter family of Hamiltonians that produce the same generating functions.

\section{Two-step walk and exclusion statistics}

The calculation of the secular determinant $F_k (\za,q)$ and of $G_{k,mn} (\za,q)$ can be most intuitively and conveniently
performed through the connection of the random walk process with exclusion statistics, as was pointed out in \cite{emeis}.

Specifically, the secular determinant $\det (1-{\bf M})$ of a paradiagonal matrix $\bf M$
with zero diagonal and two nonzero off-diagonals, one (with elements $f_n$) just above the diagonal and the other
(with elements $g_n$) $g-1$ steps below the diagonal,
is given by the {\it grand partition function} of noninteracting particles
of exclusion statistics $g$ with single-particle statistical factors $s(n) \bb =\bb e^{-\beta (\varepsilon_n-\mu)}$,
$n=0,1,2,\dots$ ($1/\beta = k_{_B} T$) given by
\be
s (n) = -g_n \, f_n\,  f_{n+1} \cdots f_{n+g-2}
\label{ex2}\ee
The spectral function $s(n)$ encodes the single-particle energy spectrum Boltzmann factor $e^{-\beta \varepsilon_n}$
together with the fugacity parameter $x=e^{\beta \mu}$. Exclusion $g$ means that
no more than one particle can occupy any set of $g$ adjacent single-particle states.

The Dyck path Hamiltonian is of the above form. However, the Motzkin path matrix $H_k$ in (\ref{hamk}) is actually {\it not} of this form, since it has a nonvanishing diagonal. Nevertheless,
it can be expressed as the grand partition function of a $g=2$ exclusion statistics system. This is achieved by realizing Motzkin paths
as two-step Dyck paths and calculating the generating function of the two-step process.

\subsection{Two-step Dyck process}

Consider a path realized by the alternation of two Dyck processes (see fig.~\ref{2steps}): one, starting at even sites $(i,2j)$
and jumping to odd sites, up to
$(i+1,2j+1)$ with weight $z_1$ or down to $(i+1,2j-1)$ with weight $z_2$, and another,
starting at odd sites $(i,2j+1)$ and jumping to even sites, up to  $(i+1,2j+2)$ with weight $z_2$ or down to
$(i+1,2j)$ with weight $z_1$.
The full process is height-restricted with floor $j=0$ and ceiling $j=2k+2$.
The $(2k+3)$-dimensional Hamiltonian transition matrix of the full process is
\be
{H}_{2D} = 
\begin{pmatrix}
0\;\;& z_1 \;\, & 0\;\; & 0\;\; & 0\;\; \cdots & 0\;\; & 0\; \\
z_1 \;\;& 0 \;\;&z_2 \;\;& 0\;\; & 0\;\; \cdots & 0 \;\;& 0\; \\
0 \;\;& z_2 \;\; & 0\;\; &z_1 q_o &0 \;\;\cdots & 0 \;\;& 0\; \\
0 \;\;& 0 \;\; &z_1 q_o &0 \;\; & z_2 q_o \cdots  & 0 \;\;& 0\; \\
0 \;\;& 0 \;\; &0\;\; &z_2 q_o & 0\;\; \cdots  & 0 \;\;& 0\; \\
\vdots \;\;& \vdots \;\;& \vdots\;\; &\vdots\;\; &\; \;\;\ddots & \vdots\;\; & \vdots\; \\
0\;\; & 0 \;\;& 0 \;\; &0\;\; &0\;\; \cdots & 0\; & z_2 q_o^k \;\\
0\;\;& 0 \;\;& 0 \;\; &0 \;\; &0 \;\;\cdots & z_2 q_o^k & 0\; \\
\end{pmatrix}
\label{2D}\ee
with $q_o$ a new area-counting parameter. Denoting the basis states of this $(2k\bb+\bb 3)$-dimensional Hilbert space
$\Cet{j}$, $j=0,1,\dots,2k+2$, $H_{2D}$ acts on them as
\bea
\Dra{2j} H_{2D} &=& z_1 q_o^j \Dra{2j\bb+\bb 1} + z_2  q_o^{j-1} \Dra{2j\bb-\bb 1} \nonumber\\
\Dra{2j\bb+\bb 1} H_{2D} &=& z_2 \, q_o^j \Dra{2j+2} + z_1 q_o^j \Dra{2j}
\eea
with $\Cet{-1} \equiv 0 \equiv \Cet{2k+3}$.

{\begin{figure}\label{2steps}\vskip -1cm
{\hskip -1.4cm
\begin{tikzpicture}[scale=0.6]

\draw[help lines, gray] (0,-0.02) grid (24.5,8.01);

\tikzset{big arrow/.style={decoration={markings,mark=at position 1 with {\arrow[scale=3,#1,>=stealth]{>}}},postaction={decorate},},big arrow/.default=black}

\draw[ultra thick,-] (0,0) -- (24.8,0);
\draw[ultra thick,-] (0,-0.04) -- (0,9);
\draw[ultra thick,-] (0,8) node[left] {$2k\bb+\bb2\bb=\bb 8$} -- (24.5,8);
\draw[line width=0.45mm,blue,-] (0,1) -- (24.5,1);\draw[line width=0.2mm,-,blue] (0,3) -- (24.5,3);
\draw[line width=0.2mm,-,blue] (0,5) -- (24.5,5);\draw[line width=0.38mm,-,blue] (0,7) node[left] {\color{black} $\ket{3}$} -- (24.5,7);
\draw[line width=0.2mm,-,blue] (2,-0.02) -- (2,8);\draw[line width=0.2mm,-,blue] (4,-0.02) -- (4,8);\draw[line width=0.2mm,-,blue] (6,-0.02) -- (6,8);\draw[line width=0.2mm,-,blue] (8,-0.02) -- (8,8);
\draw[line width=0.2mm,-,blue] (10,-0.02) -- (10,8);\draw[line width=0.2mm,-,blue] (12,-0.02) -- (12,8);\draw[line width=0.2mm,-,blue] (14,-0.02) -- (14,8);\draw[line width=0.2mm,-,blue] (16,-0.02) -- (16,8);
\draw[line width=0.2mm,-,blue] (20,-0.02) -- (20,8);\draw[line width=0.2mm,-,blue] (22,-0.02) -- (22,8);\draw[line width=0.2mm,-,blue] (24,-0.02) -- (24,8);
\draw[line width=0.2mm,-,blue] (18,-0.02) -- (18,8);

\draw[big arrow] (0,0) -- (25,0) node[below right] {$i$};

\draw[big arrow] (0,-0.02) -- (0,9) node[left] {$j$};

\draw[thick,black!30!green,-](0,1)node[left] {{\color{black} $\ket{0}$}}--(1,0);
\draw[line width=0.8mm,black!30!green,-](2,1)--(3,2); \draw[thick,black!30!green,-] (4,3)--(5,4);
\draw[line width=0.8mm,black!30!green,-] (6,3)--(7,4); \draw[thick,black!30!green,-](8,5)--(9,4);
\draw[line width=0.8mm,black!30!green,-](10,5)--(11,6); \draw[thick,black!30!green,-](12,7)--(13,8);
\draw[line width=0.8mm,black!30!green,-](14,7)--(15,6);\draw[line width=0.8mm,black!30!green,-](16,5)--(17,6);
\draw[line width=0.8mm,black!30!green,-](18,7)--(19,6);\draw[thick,black!30!green,-](20,5)--(21,6);
\draw[line width=0.8mm,black!30!green,-](22,5)--(23,4);
\draw[thick,purple,-](1,0)--(2,1); \draw[line width=0.8mm,purple,-](3,2)--(4,3); \draw[thick,purple,-](5,4)--(6,3);
\draw[line width=0.8mm,purple,-](7,4)--(8,5); \draw[thick,purple,-](9,4)--(10,5);\draw[line width=0.8mm,purple,-](11,6)--(12,7);
\draw[thick,purple,-](13,8)--(14,7); \draw[line width=0.8mm,purple,-](15,6)--(16,5);\draw[line width=0.8mm,purple,-](17,6)--(18,7);
\draw[line width=0.8mm,purple,-](19,6)--(20,5);\draw[thick,purple,-](21,6)--(22,5);
\draw[line width=0.8mm,purple,-](23,4)--(24,3)node[right] {\color{black} $~\ket{1}$};
\draw[line width=0.8mm,blue,-](0,1)--(2,1); 
\draw[line width=0.8mm,blue,-](4,3)--(6,3); 
\draw[line width=0.8mm,blue,-](8,5)--(10,5);
\draw[line width=0.8mm,blue,-](12,7)--(14,7);\draw[line width=0.8mm,blue,-](20,5)--(22,5);

\fill (0,1) circle (4pt);
\fill (24,3) circle (4pt);


\end{tikzpicture}
}
\vskip -0.8cm
\caption{\small{The two-step Dyck process: (green) steps $(2i,2j\bb+\bb 1)\to (2i\bb+\bb 1,2j\bb+\bb 1\pm1)$ with amplitudes $z_1,z_2$
alternating with (red) steps $(2i\bb+\bb1,2j) \to (2i\bb+\bb2,2j\pm1)$ with reversed amplitudes $z_2,z_1$ on a lattice with ceiling 8.
The points on the (blue) sublattice $(2i,2j+1)$ generate a Motzkin path (thick meander) of length 12 from $m=0$ to $n=1$
with ceiling $k=3$, such that $2k+2=8$.}}\label{Double Dyck}
\end{figure}
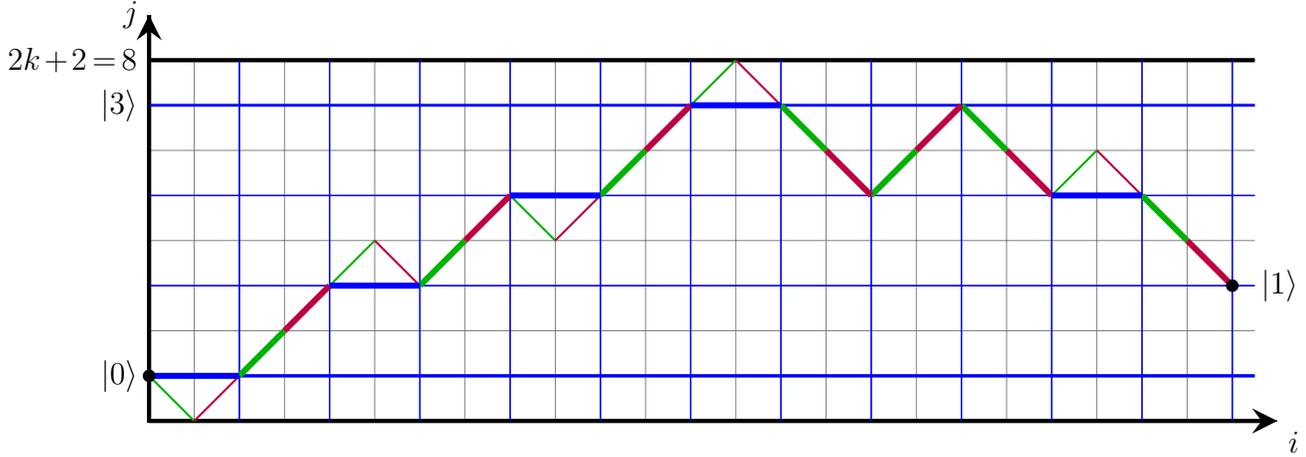}

It is clear that the path ``distilled'' from the above process by considering the height of the walk only at the odd sites
$\Cet{2j+1}$, $j=0,1,\dots,k$ every second time step is a Motzkin path of restricted height $k$ (see fig.~\ref{2steps}). Specifically, acting
with $H_{2D}$ twice on an odd state,
\be
\Dra{2j+1} H_{2D}^2 = z_1 z_2\, q_o^{2j+1} \Dra{2j+3} + (z_1^2 + z_2^2 )q_o^{2j} \Dra{2j+1}
+ z_1 z_2\, q_o^{2j-1} \Dra{2j-1}
\ee
Defining $\ket{j} = \Cet{2j+1}$, $j=0,1,\dots,k$, the above relation implies that $H_{2D}^2$ acts on the $(k+1)$-dimensional subspace
$\ket{j}$ as the Motzkin Hamiltonian
\be
\bra{j}  H_{2D}^2 = z q^{j+1/2} \bra{j+1} + z_h q^j \bra{j} + z q^{j-1/2} \bra{j-1}
\label{oddM}\ee
provided we identify
\be
q = q_o^2 ~,~~ z = z_1 z_2 ~,~~ z_h = z_1^2 + z_2^2 ~~~\Rightarrow~~~ z_{1,2}^2 = {z_h \over 2} \pm \sqrt{{z_h^2\over 4} - z^2}
\label{map}\ee
(Note that the choice of roots for $z_1$ and $z_2$ in (\ref{map}), or equivalently the order of the two
Dyck processes, is irrelevant; the Motzkin process does not depend on that choice.)
We will also adopt the $(z,\omega)$ parametrization
\be
z_1^2 = z\omega ,~ z_2^2 = z \omega^{-1},~ \omega = {z_h \over 2z}+ \sqrt{{z_h^2 \over 4z^2}- 1}
~~ \Rightarrow~~ z_h = z ( \omega + \omega^{-1})
\ee
so that powers of $z$ count the total number of steps: $z^{l_u} z_h^{l_h} z^{l_d} = z^{l_u+l_h+l_d}
(\omega + \omega^{-1} )^{l_h}$.
 
We note that the complementary process restricted to even states also generates a Motzkin process, of height $k+1$.
The amplitudes of the steps $\Cet{0} \to \Cet{0}$ and $\Cet{2k+2} \to \Cet{2k+2}$, however, are truncated,
since the intermediate steps $\Cet{0} \to \Cet{-1}$ and $\Cet{2k+2} \to \Cet{2k+3}$
are missing and do not contribute to the amplitude. The odd states, on the other hand, provide a faithful realization
of Motzkin paths with the proper weights, upon identifying states and parameters as in (\ref{oddM}) and (\ref{map}).

It remains to connect the secular determinant of the Motzkin process with that of the two-step Dyck process.
To this end, define the projection operator on odd states $P$ and the projector on even states ${\bar P} = 1-P$. Clearly
\be
H_{2D} P = {\bar P} H_{2D} ~,~~ H_{2D} {\bar P} = P H_{2D} ~,~~ P\bar P = \bar P P = 0~,~~ P+\bar P = 1
\label{PP}\ee
The Motzkin Hamiltonian $H_k$ and the ``complementary'' quasi-Motzkin Hamiltonian ${\bar H}_{k+1}$ are,
up to zero modes when acting on the `wrong' subspace,
\be
H_k = H_{2D}^2 P ~,~~~ {\bar H}_{k+1} = H_{2D}^2 \bar P
\ee
and therefore
\bea
\det (1-H_k) &=& \det (1- H_{2D}^2 P ) = \det (1- H_{2D} {\bar P} H_{2D} P ) \nonumber \\
&=& \det (1- H_{2D} P H_{2D} {\bar P} ) = \det(1- {\bar H}_{k+1})
\eea
The zero modes are irrelevant, due to the presence of the unit matrix, so the secular determinants of the Motzkin
process and its complement in their respective subspaces are equal. Thus
\bea
\det(1-H_{2D}^2) &=& \det \left[ (1-H_{2D}^2)P + (1-H_{2D}^2){\bar P} \right] \nonumber\\
&=& \det(1-H_k) \det (1-{\bar H}_{k+1} ) = \det(1-H_k)^2
\label{H2H}\eea
Finally, (\ref{PP}) implies that $H_{2D}$ anticommutes with the parity matrix $\Sigma \bb=\bb {\bar P}\bb -\bb P,\, 
\Sigma^2 \bb=\bb1$.
Therefore,
\be
\det(1-H_{2D}) = \det (1 + \Sigma H_{2D} \Sigma) = \det[\Sigma (1+H_{2D}) \Sigma] = \det(1+H_{2D})
\ee
and
\be
\det(1-H_{2D}^2) = \det(1-H_{2D}) \det(1+H_{2D}) = \det(1-H_{2D})^2
\ee
Comparing with (\ref{H2H}) we eventually obtain
\be
\det(1-H_{2D}) = \det(1-H_k )
\ee
(the sign is fixed by continuity from $\za = 0$).

The end result is that we can calculate the secular determinant
$F_k (\za,q)$ by evaluating instead the secular determinant of the two-step Dyck path Hamiltonian.

\subsection{Two-step secular determinant and bosonization}

The two-step Hamiltonian is of the $g=2$ exclusion statistics form. The spectral
parameters can be read off from the product of conjugate off-diagonal elements:
\be
s(2n) = -z_1^2 \, q_o^{2n} \, ,~~~ s(2n+1) = -z_2^2 \, q_o^{2n} \,
,~~~ n=0,1,\dots,k
\label{speca}\ee
Calling, for ease of distinction, $s(2n) = \alpha(n)$, $s(2n+1) = \beta (n)$ and remembering that $z_{1,2}^2 = 
z \omega^{\pm 1}$, $q_o^2 = q$, the spectral parameters (\ref{speca}) are
\be
\alpha(n) = -z\omega \, q^n,~~ \beta(n) = -z\omega^{-1} q^n,~ n = 0,1,\dots,k
\label{alphabeta}\ee
The secular determinant $F_k (\za,q)$ is the grand partition function of exclusion-2 particles in
levels with spectral parameters $s(0),s(1),\dots,s(2k+1)$ in that order.
Particles placed on these levels must have at least one empty level between them. Calling $Z_{k,N}$ the
$N$-body partition function in the above spectrum, the grand partition function ${\cal Z}_k$ becomes
\be
{\cal Z}_k = \sum_{N=0}^{k+1} Z_{k,N} = \sum_{N=0}^{k+1}\, 
\sum_{\{0\le n_i \le n_{i+1} -2 \le 2k-1\}} s(n_1) s(n_2) \cdots s(n_N)
\ee
$n_i = 0,1,\dots,2k+1$ mark the levels on which particles are placed, in increasing order, and the condition
$n_i \le n_{i+1} -2$ in the partition function $Z_N$ enforces exclusion-2 statistics. It is clear that at most $k+1$
particles can be accommodated in the available $2k+2$ levels.

In the case of Dyck paths, the spectral factors $s(n)$ corresponded to the equidistant spectrum of a truncated
harmonic oscillator and the partition function could be found using bosonization.
For general exclusion statistics $g$, bosonization is achieved by redefining the occupied level numbers 
$n_i \le n_{i+1}\bb-\bb g$ as
\be
n_i = \ell_i +g(i-1) ~~~ \Rightarrow ~~~ \ell_i \le \ell_{i+1}
\label{nl}\ee
This reduces the ``gap'' between successive occupied levels $\ell_i$ and $\ell_{i+1}$ by $g$, making the new
occupation numbers $\ell_i$ obey bosonic statistics. 
The $N$-body spectral factor becomes
\be
s(n_1) s(n_2) \cdots s(n_N) = s(\ell_1) s(\ell_2 +g) \cdots s(\ell_N + (N-1)g)
\label{snl}\ee
In general, this is no simpler than the expression in terms of $n_i$. Bosonization becomes useful for an equidistant spectrum,
for which $s(n) = a q^n$ for some $a,q$. In that case (\ref{snl}) gives
\be
s(\ell_1) s(\ell_2 +g) \cdots s(\ell_N + (N-1)g) = q^{g {N(N-1) \over 2}} s(\ell_1) s(\ell_2) \cdots s(\ell_N)
\ee
and the $N$-body exclusion-$g$ partition function $Z_{N}^{(g)}$ becomes the bosonic one $Z_{N}^{(B)}$
up to an overall coefficient
\be
Z_{N}^{(g)} = q^{g {N(N-1) \over 2}} Z_{N}^{(B)}
\label{gB}\ee

{\begin{figure} \vskip -1.7cm
{\hskip 2.4cm
\begin{tikzpicture}[scale=0.36]

\tikzset{big arrow/.style={decoration={markings,mark=at position 1 with {\arrow[scale=3,#1,>=stealth]{>}}},postaction={decorate},},big arrow/.default=black}

\draw[ultra thick,black!30!green,-](0,0) node[left]{$\alpha(0)$} -- (2,0);
\draw[ultra thick,black!30!green,-](0,2) node[left]{$\alpha(1)$} -- (2,2);
\draw[ultra thick,black!30!green,-](0,4) -- (2,4);
\draw[ultra thick,black!30!green,-](0,6) -- (2,6);
\draw[ultra thick,black!30!green,-](0,8) -- (2,8);
\draw[ultra thick,black!30!green,-](0,10) -- (2,10);
\draw[ultra thick,black!30!green,-](0,12) -- (2,12);
\draw[ultra thick,black!30!green,-](0,14) -- (2,14);
\draw[ultra thick,black!30!green,-](0,16) -- (2,16);
\draw[ultra thick,black!30!green,-](0,18) node[left]{$\alpha(9)$}-- (2,18);
\draw[ultra thick,black!30!green,-](0,20) -- (2,20);

\draw[ultra thick,blue,-](2,0.7) -- (4,0.7);
\draw[ultra thick,blue,-](2,2.7) -- (4,2.7);
\draw[ultra thick,blue,-](2,4.7) -- (4,4.7)node[right]{$\beta(2)$};
\draw[ultra thick,blue,-](2,6.7) -- (4,6.7);
\draw[ultra thick,blue,-](2,8.7) -- (4,8.7);
\draw[ultra thick,blue,-](2,10.7) -- (4,10.7)node[right]{$\beta(5)$};
\draw[ultra thick,blue,-](2,12.7) -- (4,12.7)node[right]{$\beta(6)$};
\draw[ultra thick,blue,-](2,14.7) -- (4,14.7)node[right]{$\beta(7)$};
\draw[ultra thick,blue,-](2,16.7) -- (4,16.7);
\draw[ultra thick,blue,-](2,18.7) -- (4,18.7);
\draw[ultra thick,blue,-](2,20.7) -- (4,20.7) node[right]{$\beta(10)$};


\draw[ultra thick,black!30!green,-](20,0) node[left]{$\alpha(0)$} -- (22,0);
\draw[ultra thick,black!30!green,-](20,2) node[left]{$\alpha(1)$} -- (22,2);
\draw[ultra thick,black!30!green,-](20,4) node[left]{$\alpha(2)$} -- (22,4);
\draw[ultra thick,black!30!green,-](20,6) node[left]{$\alpha(3)$} -- (22,6);

\draw[ultra thick,blue,-](22,0.7) -- (24,0.7)node[right]{$\beta(0)$};
\draw[ultra thick,blue,-](22,2.7) -- (24,2.7)node[right]{$\beta(1)$};
\draw[ultra thick,blue,-](22,4.7) -- (24,4.7)node[right]{$\beta(2)$};
\draw[ultra thick,blue,-](22,6.7) -- (24,6.7)node[right]{$\beta(3)$};

\fill[purple] (20.7,0) circle (5pt);
\fill[purple] (21.3,0) circle (5pt);
\fill[purple] (23,0.7) circle (5pt);
\fill[purple] (23,6.7) circle (5pt);
\fill[purple] (22.5,4.7) circle (5pt);
\fill[purple] (21,6) circle (5pt);
\fill[purple] (23.5,4.7) circle (5pt);
\fill[purple] (23,4.7) circle (5pt);

\fill[purple] (1,0) circle (5pt);
\fill[purple] (1,2) circle (5pt);
\fill[purple] (3,4.7) circle (5pt);
\fill[purple] (3,12.7) circle (5pt);
\fill[purple] (3,10.7) circle (5pt);
\fill[purple] (1,18) circle (5pt);
\fill[purple] (3,14.7) circle (5pt);
\fill[purple] (3,20.7) circle (5pt);

\fill (12,6) circle (0.1pt) node[above]{\small{\bf Bosonization}};
\fill (1.8,-1) circle (0.1pt) node[below]{\small{Exclusion-$2$ particles}};
\fill (22,-1) circle (0.1pt) node[below]{\small{Bosons}};
\draw[ultra thick] (10.5,5) -- (13.5,5);
\draw[thick,big arrow] (13.5,5) -- (13.7,5);

\end{tikzpicture}
}\vskip -0.1cm
\caption{\footnotesize{An example of bosonization for a state of $g=2$, $k=10$, $N=8$. In the exclusion-2 picture,
on the left, there are
11 (green) $\alpha$-levels and 11 (blue) $\beta$-levels. The distance between successive $\alpha$ or $\beta$ levels
represents a factor of $q$, while the distance between an $\alpha$ level and the next $\beta$ level
represents a factor of $\omega^{-2}$. 
Particles, represented by (red) dots, cannot occupy the same or neighboring levels, irrespective of level type. Bosonization shrinks $k$ to $k-(N-1) = 3$ and lowers the $i^{\text{th}}$
lowest particle by $i-1$ steps in its own level type: $\alpha(0) \to \alpha(0), \alpha(1) \to \alpha(1-1) = \alpha(0),
\beta(2) \to \beta(2-2) = \beta(0), \beta(5) \to \beta(5-3) = \beta(2)$ and so on, leading to the bosonic state on the right.
The number of particles in $\alpha$-levels ($N_\alpha =3$) and $\beta$-levels ($N_\beta = 5$) is preserved.
The total ``height'' of particles has been reduced by $N(N-1)/2 = 28$ leading to a factor of $q^{N(N-1)/2} = q^{28}$
relating the $g=2$ state to the bosonic state.}}\label{Bosonization}
\end{figure}
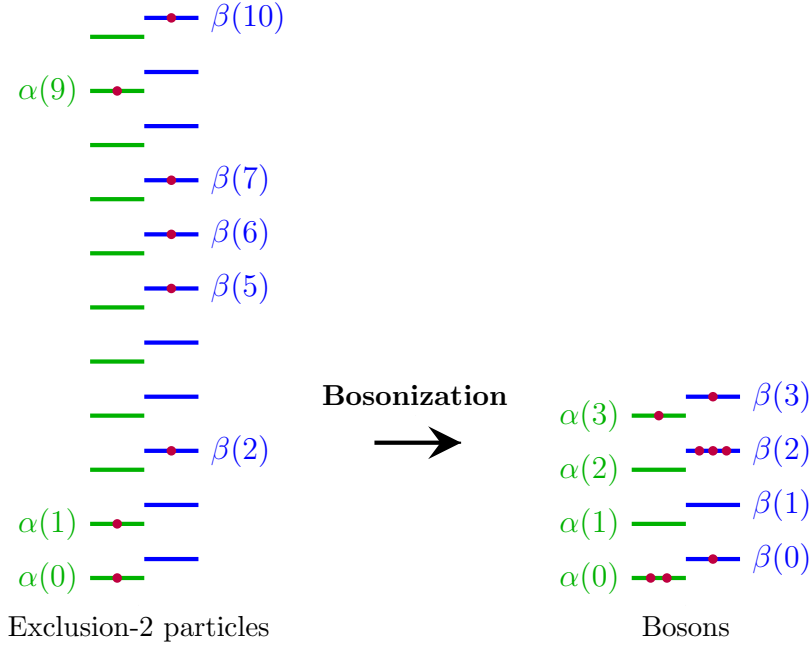}

Remarkably, bosonization
also works for the two-step spectrum of the Motzkin process, although the spectrum corresponding to $s(n)$ is not equidistant.
The redefinition (\ref{nl}) for the two-step Dyck (i.e., Motzkin) case, with $g=2$, involves a shift by an {\it even} integer, so it
maps $\alpha$-levels to $\alpha$-levels and $\beta$-levels to $\beta$-levels (see fig.\ref{Bosonization}):
\bea
\hskip -1.5cm n_i = \ell_i\bb +2(i\bb-\bb 1) ~~ \text{so:}~~~
\ell_i &=& 2 l_i\bb :~~~~~~\, s(n_i )= \alpha(l_i\bb+\bb i\bb -\bb1) =q^{i-1} \alpha(l_i) = q^{i-1} s(\ell_i)\nonumber\\
\ell_i &=& 2 l_i \bb+\bb 1\bb :~~ s(n_i)= \beta(l_i\bb+\bb i\bb -\bb1) =q^{i-1} \beta(l_i) = q^{i-1} s(\ell_i)
\eea
Consequently, $s(n_i) = q^{i-1} s(\ell_i)$
and the relation (\ref{gB}) remains valid. The range of $\ell_i$, though, is reduced to $0,1, \dots, 2k+1-2(N-1)$. Therefore,
\be
{\cal Z}_k = \sum_{N=0}^{k+1} Z_{k,N} = \sum_{N=0}^{k+1} q^{N(N-1)\over 2} Z_{k-N+1,N}^{(\alpha \beta)}
\ee
$Z_{k-N+1,N}^{(\alpha \beta)}$ is the partition function of bosons distributed to two towers of equidistant levels,
$\alpha(n)$ and $\beta(n)$,
with level numbers in each $l_i = 0,1,\dots,k\bb-\bb N+1$. The partition function for fixed numbers of particles $N_\alpha$ and
$N_\beta$ in each tower factorizes. The full grand
partition function, however, does not, due to the factor $q^{N(N-1)/2}$ that involves $N_\alpha + N_\beta = N$, and the fact
that each tower contains $k-N_\alpha-N_\beta+1$ levels, coupling the two towers.

\subsection{Calculation of the secular determinant}

We now have all the components for calculating ${\cal Z}_k = \det(1-H_{2D}) = F_k (\za,q)$. The bosonic partition function
of $N$ particles in a truncated equidistant spectrum $s(n) = a q^n$, $n=0,1,\dots,k$ is (see \cite{DyckPO} for other
alternative expressions)
\be
Z_{k,N}^{(\bb B\hskip -0.2mm )} = a^N \prod_{j=1}^N {1-q^{j+k} \over 1-q^j} = a^N \prod_{j=1}^k {1-q^{j+N} \over 1-q^j}
\label{bosonic}\ee
Applying the above to towers of type $\alpha$ or $\beta$ with spectra as in (\ref{alphabeta}), for $N_\alpha$ and $N_\beta$
particles and $k \to k-N+1$, $N=N_\alpha+N_\beta$, we have
\bea
Z_{k-\bb N_\alpha\bb-\bb N_\beta+1,N_\alpha}^{(\alpha)}\bb\bb &=& (-z\omega)^{N_\alpha}\prod_{j=1}^{N_\alpha} {1-q^{j+k-\bb N_\alpha\bb-\bb N_\beta+1} \over 1- q^j}
= (-z\omega)^{N_\alpha}\bb \prod_{j=1}^{k-\bb N_\alpha\bb -\bb N_\beta+1} {1-q^{j+N_\alpha} \over 1-q^j}
\hskip 1.7cm {} \label{ZaZb}\\
Z_{k\bb-N_\alpha\bb-\bb N_\beta+1,N_\beta}^{(\beta)}\bb\bb &=& (-z\omega^{-1})^{N_\beta}\prod_{j=1}^{N_\beta} {1-q^{j+k-N_\alpha-N_\beta+1} \over 1- q^j}
= (-z\omega^{-1})^{N_\beta}\bb \prod_{j=1}^{k-\bb N_\alpha\bb-\bb N_\beta+1} {1-q^{j+N_\beta} \over 1-q^j} \nonumber
\eea
and
\be
{\cal Z}_k = \sum_{N_\alpha,N_\beta=0}^{N_\alpha+N_\beta \le k+1} q^{(N_\alpha + N_\beta)(N_\alpha+N_\beta-1)\over 2}
Z_{k-\bb N_\alpha\bb-\bb N_\beta+1,N_\alpha}^{(\alpha)}\, Z_{k-\bb N_\alpha\bb-\bb N_\beta+1,N_\beta}^{(\beta)}
\ee
Using either the first or the second expressions in (\ref{ZaZb}) and changing summation variables
we obtain for ${\cal Z}_k = F_k(\za,q)$ the two alternative forms
\bea
F_k(\za,q) &=&\bb \sum_{N=0}^{k+1}  (-z)^{N} \sum_{n=0}^{N}
\omega^{N-2n}
\prod_{j=1}^n {q^{N-1 \over 2} - q^{j+k-{N-1 \over 2}} \over 1-q^j}\,
\prod_{l=1}^{N-n} {q^{N-1 \over 2} - q^{l+k-{N-1 \over 2}} \over 1-q^l} ~~~{}
\nonumber\\
&=& \bb \sum_{N=0}^{k+1}  (-z)^{N} q^{N(N-1)\over 2}\sum_{n=0}^{N}
\omega^{N-2n}
\prod_{j=1}^{k-N+1} {(1-q^{j+n} )(1-q^{j+N-n} ) \over (1-q^j )^2}
\label{wow1}\eea
Although 
$\omega$ can be complex, the above expressions are invariant under 
$\omega \to \omega^{-1} = \omega^*$
(upon $n \to N-n$, for real $z$) and thus are real. In fact, we can use this property to express $F_k(\za,q)$ in terms of
Chebyshev polynomials $T_n (z_h/2z)$. Adding the expressions for $\omega$ and $\omega^{-1}$ in (\ref{wow1},\ref{wow2})
we obtain
\bea
F_k(\za,q) &=&\bb \sum_{N=0}^{k+1}  (-z)^{N} \sum_{n=0}^{N}
T_{|2n-N|}\bb\bb \left({z_h \over 2 z}\right)
\prod_{j=1}^n {q^{N-1 \over 2} - q^{j+k-{N-1 \over 2}} \over 1-q^j}\,
\prod_{l=1}^{N-n} {q^{N-1 \over 2} - q^{l+k-{N-1 \over 2}} \over 1-q^l} ~~~{}\nonumber\\
&=&\bb \sum_{N=0}^{k+1}  (-z)^{N} q^{N(N-1)\over 2}\sum_{n=0}^{N}
T_{|2n-N|}\bb\bb \left({z_h \over 2 z}\right)
\prod_{j=1}^{k-N+1} {(1-q^{j+n} )(1-q^{j+N-n} ) \over (1-q^j )^2}
\label{wow2}\eea
Eqs.~(\ref{wow1},\ref{wow2}) are our third main result. Using the above expressions for $F_k(\za,q)$ in (\ref{typara}),
we obtain the generating function of Motzkin paths $G_{k,mn} (\za,q)$.

Expressions (\ref{wow2}) are explicit polynomials in $z,z_h$. For even $N$, only even Chebyshev polynomials appear
with degrees $0$ up to $N$, leading to polynomials in $z,z_h$ of total degree $N$ and terms from $z_h^N$ to $z^N$.
For odd $N$ only odd Chebyshev polynomials appear, leading again to total degree $N$ polynomials in $z,z_h$ but now
with terms from $z_h^N$ to $z^{N-1} z_h$. This is related to the fact that excursions with an odd number of steps
cannot consist entirely of up and down steps (factors $z$) and must contain at least one horizontal step ($z_h$).

Finally, we point out that the products in the above formulae, as well as in other
formulae in this paper, can be expressed in terms of the
$q$-Pochhammer symbol $(a;q)_n$, but we will not do this transcription.

\subsection{Checks and special cases}

A few checks can be performed on formulae (\ref{wow1},\ref{wow2}), reducing them to known cases.

\noindent
{\bf a}. Identical underlying Dyck processes: $ z_1 \bb=\bb z_{_D} q_{_D}^{1/2}$, $z_2 \bb=\bb z_{_D} q_{_D}^{3/2}$,
$q_o = q_D^2$.
This reduces the two-step process (\ref{2D}) into a standard Dyck path process with parameters $z_{_D},q_{_D}$.
From (\ref{map},\ref{speca}) the above choices imply
\be
q=q_{_D}^4 ~,~z = z_{_D}^2 q_{_D}^2 ~,~ z_h = z_{_D}^2 (q_{_D}+q_{_D}^3) ~,~ \omega = q_{_D} ~; ~~~
s(n) = -z_{_D}^2 q_{D}^{2n+1} 
\ee
$s(n)$ is the standard equidistant spectrum of the Dyck process. (\ref{wow1}) reproduces the Dyck path determinant
given in \cite{DyckPO} upon use of the identity
\be
\sum_{n=0}^N q^n \prod_{j=1}^{k} 
{\left(1-q^{2(j+n)}\right)\left(1-q^{2(j+N-n)}\right) \over (1-q^{2j})^2} = \prod_{j=1}^{2k+1} {1-q^{j+N} \over 1-q^j}
\ee

\noindent
{\bf b}. $q=1$: the generating function accounts only for length and type of steps. $F_k (z_i,1)$ degenerates to
\be
F_k(\za,1) =\bb \sum_{N=0}^{k+1}  (-z\omega)^{N} \sum_{n=0}^{N}
\omega^{-2n} {k+1-N+n \choose n}{k+1-n \choose N-n}
\ee

\noindent
{\bf c}. $z_h =0$: the horizontal step is suppressed and the process degenerates into Dyck paths. (\ref{alphabeta}) implies
$\omega = i$ and $\alpha(n) = -\beta(n) = -iz q^n$. This actually eliminates all odd $N$ from the sum in
(\ref{wow1},\ref{wow2}) and reproduces the Dyck path determinant, as we shall demonstrate in the next subsection.

\noindent
{\bf d}. $z_h = z$, weighting all steps equally. (\ref{alphabeta}) implies
\be
\omega = {1\over 2} +i{\sqrt 3 \over 2} = e^{i\pi /3}
\ee 
In this case too the determinant assumes a special form.

Cases c.~and d.~are instances of a subset of values of $z,z_h$ for which the grand partition function admits a special
interpretation and the determinant has special properties. We treat these special cases in the next subsection.

\noindent
{\bf e.} $k=\infty$ (no ceiling). In this case the formula simplifies to
\be
F_k(\za,q) =\bb \sum_{N=0}^{k+1}  (-z)^{N} q^{N(N-1)\over 2}\sum_{n=0}^{N}
\prod_{j=1}^n {\omega \over 1-q^j}\,
\prod_{l=1}^{N-n} {\omega^{-1} \over 1-q^l} ~~~{}
\label{wowinf}\ee 

\subsection{A ``dual'' form of the determinant and cyclic cases}

An alternative form for the secular determinant $F_k(\za,q)$ can be obtained by considering the bosonized system, instead of
two ``vertical'' towers of levels, $\alpha(n)$ and $\beta(n)$ with $k-N+2$ levels each, as $k-N+2$ ``horizontal''
sets of two levels each.
Calling $N_j$ the number of particles in set $j$ with levels $\alpha(j),\beta(j)$ ($j=0,1,\dots,k-N+1$) the $N_j$-particle
bosonic partition function for set $j$ is
\be
\hskip -0.1cm {Z}_{j;N_j}^{(\bb B\hskip -0.2mm )} \bb=\bb \sum_{n=0}^{N_j} \alpha(j)^n \beta(j)^{N_j -n}= (-z\omega^{-1})^{N_j} q^{jN_j} \sum_{n=0}^{N_j} \omega^{2n}
\bb=\bb (-z\omega^{-1})^{N_j} q^{jN_j} {1-\omega^{2(N_j+1)} \over 1-\omega^2}
\ee
Accounting for the factor $q^{N(N-1)/2}$ relating the bosonic to the exclusion-2 partition function, the grand partition
function (secular determinant) becomes
\be
F_k(\za,q) =\bb \sum_{N=0}^{k+1}  (-z\omega^{-1})^{N} q^{N(N-1)\over 2} \sum_{\left\{ \sum N_j=N\right\}}
\prod_{j=0}^{k-N+1}q^{jN_j}{1-\omega^{2(N_j+1)} \over 1-\omega^2}~~~{}
\label{wow3}\ee 
This form may look less useful that (\ref{wow1}) or (\ref{wow2}), as it involves multiple sums, but is better in revealing the
structure of the special systems we will study in the sequel. Enforcing the constraint
$\delta ( \sum_j N_j -N)$ in terms of an exponential integral, and harmlessly extending the summation range 
of the $N_j$ to infinity, the above can also be rewritten as
\be
F_k(\za,q) =\bb \int_0^{2\pi} \bb\bb d\theta \,\sum_{N=0}^{k+1}\,
 {  \left(-z e^{-i\theta}\right)^{N}q^{N(N-1)\over 2}} \prod_{j=0}^{k-N+1}{1 \over (1-\omega^{-1} e^{i\theta} q^j)(1-\omega\, e^{i\theta} q^j)}~~~{}
\label{wow4}\ee 

We now focus our attention to ``cyclic'' walks with parameters such that $\omega^2$ is a root of unity; that is,
\be
\omega = e^{i \pi p /r} ~~~ \text{or} ~~~ z_h = 2z \cos{\pi p \over r}~,~~~ r = 1,2,\dots ~,~~p,r ~\text{coprime}
\ee
$p\bb=\bb 1,r\bb=\bb 2$ corresponds to case c.~of the previous subsection, while $p\bb=\bb 1,r\bb=\bb 3$ corresponds to case  d. For such values of $\omega^2$ the $\omega$-dependent ratio inside the product in (\ref{wow3})
is periodic in $N_j$ with period $r$ and vanishes for $N_j = -1 (\text{mod}~ r)$.

To capitalize on this property we put
$N_j = r n_j + \ell_j$, $n_j = 0,1,\dots$, $\ell_j = 0,1,\dots,r-2$; (\ref{wow3}) becomes
\be
F_k(\za,q) = \bb \sum_{N=0}^{k+1}  (-z\omega^{-1})^{N} q^{N(N-1)\over 2} 
\bb\bb\sum_{\left\{\bb {r\sum n_j +\sum \ell_j =N \atop \ell_j=0,1,\dots,r-1}\bb\right\}}^N
\bb\prod_{j=0}^{k-N+1} q^{r j n_j} \prod_{j=0}^{k-N+1} q^{j \ell_j} {1-\omega^{2(\ell_j+1)} \over 1-\omega^2}~~~{} 
\label{bopa}\ee
We can interpret $n_j$ as counting bosons and $\ell_j$
as counting `parafermions' of order $r-\bb 2$ with the total number of bosons $n$ and parafermions $\ell$ satisfying
$N = r n+\ell$. By parafermions we mean particles with the property that up to $r\bb-\bb 2$ of them can be placed in a
single-particle level. Then the first product in (\ref{bopa}) is the partition function of $n$ bosons in levels $q^{rj}$ while
the second product is the corresponding parafermionic partition function. Reverting to (\ref{bosonic}) for the bosonic
partition function, and putting $N=rn+\ell$,
\be
F_k(\za,q) = \bb \sum_{n,\ell} (-z\omega^{-1})^{rn+\ell} q^{(rn+\ell)(rn+\ell-1)\over 2} 
 \, Z_{k-rn-\ell+1}^{\{r-2\}} (\ell) \prod_{j=1}^{n} {1-q^{r(j+k-rn-\ell+1)} \over 1-q^{rj}}
\label{boo}\ee
with $Z_{k-rn-\ell+1}^{\{r-2\}} (\ell)$ the parafermionic partition function for $\ell$ particles
\be
Z_{k-rn-\ell+1}^{(r-2)} (\ell) = \bb
 \sum_{\ell_j =0 \atop \left\{\sum \ell_j = \ell \right\}}^{r-2}
\prod_{j=0}^{k-rn-\ell+1} q^{j \ell_j} {1-\omega^{2(\ell_j+1)} \over 1-\omega^2}
\ee
We can now examine some special cases.

\noindent
{\bf c}. $r=2,p=1 \Rightarrow \omega=i$: only $\ell_j=0$ survives, so there are no parafermions. We get
\bea
F_k(z,z_h=0,q) &=&
\bb \sum_{n} (zi)^{2n} q^{n(2n-1)} 
\prod_{j=1}^{n} {1-q^{2(j+k-2n+1)} \over 1-q^{2j}} \nonumber\\
&=& \bb \sum_{n} (-z^2)^{n} (q^2)^{n(n-1)} 
\prod_{j=1}^{n} (q^2)^{1/2}{1-(q^2)^{j+k-2n+1} \over 1-(q^2)^{j}}
\eea
This is precisely the secular determinant of the Dyck path process calculated in \cite{DyckPO} with length
parameter $z^2$ and area parameter $q^2$, as expected.

\noindent
{\bf d}. $r\bb=\bb 3,p\bb=\bb 1 \Rightarrow \omega = e^{i\pi/3}$: here $\ell_j = 0,1$, and parafermions become ordinary fermions. The fermionic
partition function  $Z_{k-3n-\ell+1}^{(\bb F\hskip -0.2mm )} (\ell)$ can itself be bosonized. Omitting the intermediate steps, the final result is
\be
F_k(z,z,q) = \sum_{n,\ell} (-1)^n z^{3n+\ell} q^{{3n(3n-1)\over 2}+3n\ell+\ell(\ell-1)}
\prod_{j=1}^n {1-q^{3(j+k-3n-\ell+1)} \over 1-q^{3j}}
\prod_{s=1}^\ell {1-q^{s+k-3n-2\ell+2} \over 1-q^s}
\ee
(the summation in $n,\ell$ is over the values for which the summand does not vanish).
This expression is preferable to the general expression (\ref{wow1}) or (\ref{wow2}) only in that it is
manifestly real and a polynomial in $z,q$ with integer coefficients.

\section{Cluster expressions of generating functions}

In the previous section we obtained relatively explicit formulae for $F_k (\za,q)$, and therefore for $G_{k,mn} (\za,q)$.
Their form, however, is rather complicated, and their dependence on $z_h$ through $\omega$ is obscured.
In this section we will take further advantage of the connection to exclusion statistics to express the
logarithm of the generating function $\ln G_{k,mn}$ in terms of cluster coefficients. 

\subsection{Cluster coefficients}

For a grand partition function $\cal Z$, cluster coefficients $b_a$, $a=1,2,\dots$, are defined in terms of the expansion
of the grand potential $\ln {\cal Z} (x)$ in terms of the fugacity parameter $x=e^{\beta \mu}$
\be
\ln {\cal Z} (x) = \ln\left(\sum_{N=0}^{\infty}x^N Z_N  \right)=\sum_{a=1}^{\infty} x^a\, b_a
\ee
In our case, extracting the factor $x^N = (-z)^N$ out of the $N$-body partition function, $Z_{k,N} = (-z)^N {\tilde Z}_{k,N}$
[see (\ref{wow1},\ref{wow2})] as a fugacity parameter, we have the cluster expansion
\be
\ln F_k (\za,q) = \ln\left(\sum_{N=0}^{k+1} (-z)^N {\tilde Z}_{k,N} (\omega,q) \right)=\sum_{a=1}^{\infty} (-z)^a\, 
b_{k,a} (\omega,q)
\label{Fbn}\ee
The expression of the cluster coefficients for general exclusion statistics $g$ was derived in \cite{nous,emeis}.
For exclusion $g=2$, relevant to our case, with spectral parameter $s(r)$, they are expressed as a sum over all
compositions of the integer $a$ and read
\be
(-z)^a b_{k,a} = (-1)^{a-1}\hskip -0.3cm \sum_{l_1, l_2, \ldots, l_{j} ;\, j\le 2k+2\atop { \rm{compositions}}\;{\rm of}\;a}
\hskip -0.4cm 
c_2 (l_1,l_2,\ldots,l_{j} )\sum _{r=0}^{2k+2-\hskip -0.2mm j} \prod_{i=1}^{j} s^{l_{i}} (r+i-1) 
\label{bka}\ee
(Compositions are partitions where the order of terms also matters.) In our case, since the spectrum has $2k+2$ states,
only compositions with at most $2k+2$ components are possible. 
The combinatorial coefficients $c_2 (l_1,l_2,\ldots,l_{j} )$ depend only on the composition and the statistics, and for $g=2$ they are
\be
{c_2 (l_1,l_2,\ldots,l_{j})} 
= {1 \over l_1} \prod_{i=1}^{j-1} {l_i + l_{i+1}-1  \choose l_{i+1}}
= {{\prod_{i=1}^{j-1} (l_i + l_{i+1} -1)! \over \prod_{i=2}^{j-1} (l_i -1 )! \prod_{i=1}^j { l_i!}} }
\label{cl}\ee
The dependence of $b_{k,a}$ on $\za$ and $q$ is entirely through the dependence of $s(r)$ on these
parameters.

\subsection{Cluster expansion of the generating function}

The logarithm of the generating function $\ln G_{k,mn} (\za,q)$ follows from (\ref{typara}) as
\bea
\ln G_{k,mn} (\za,q) &= &(n-m) \ln z + {n^2 - m^2 \over 2} \ln q  \\
&+&\ln F_{m-1} (\za,q) + \ln F_{k-n-1} (\za q^{n+1} ,q) 
- \ln F_k (\za,q) \nonumber
\eea
for $m\le n$, and similarly for $m\ge n$. The terms in the second line are all given by (\ref{Fbn}) and (\ref{bka}),
with a common $c_2 (l_1,l_2,\dots,l_j )$, differing only in the last sums over $r$ in (\ref{bka}).
These sums for the three terms can be brought to a common form by noticing that the dependence of the
spectral factors $s(r)$ on $\za$ and $q$ implies
\be
s(q^m \za;r) = s(\za; r+2m)
\ee
as is clear from (\ref{speca}) or (\ref{alphabeta}). The three sums combine as
\bea
\ln G_{k,mn} (\za,q)
&= &(n-m) \ln z + {n^2 - m^2 \over 2} \ln q \\
+\sum_{a=1}^{\infty} (-1)^{a-1} \hskip -0.0cm 
&& \hskip -1cm \sum_{l_1, l_2, \ldots, l_{j} \atop { \rm{compositions}}\;{\rm of}\;a}
\hskip -0.6cm  c_2 (l_1,l_2,\ldots,l_{j} )
\left( \sum _{r=0}^{2m-\hskip -0.2mm j} + \sum _{r=2n+2}^{2k+2 -\hskip -0.2mm j} 
- \sum _{r=0}^{2k+2 -\hskip -0.2mm j} \right) \prod_{i=1}^{j} s^{l_{i}} (r\bb+\bb i\bb-\bb\bb 1)  \nonumber
\eea
and telescoping the sums we finally obtain
\bea
\ln G_{k,mn} (\za,q)
&= &(n-m) \ln z + {n^2 - m^2 \over 2} \ln q \label{logG} \\
&\hskip -0.8cm +& \hskip -0.45cm \sum_{a=1}^{\infty} (-1)^{a} \hskip -0.4cm \sum_{l_1, l_2, \ldots, l_{j} \atop { \rm{compositions}}\;{\rm of}\;a}
\hskip -0.4cm c_2 (l_1,l_2,\ldots,l_{j} ) \hskip 0.05cm
 \sum _{r={\max\bb{(\hb 2m+1-j,0\hskip -0.2mm)}}}^{\min\bb{(\hb2k+2-j,2n+1\hb)} } \,\prod_{i=1}^{j} s^{l_{i}} (r\bb+\bb i\bb-\bb\bb1)  \nonumber
 \eea
with the understanding that sums vanish when their lower limit exceeds their upper limit. It is clear that only compositions
with length $j$ up to $2k+2$ will contribute. Since $s(r)$ is proportional to $z$, it is clear that the above sum is
an expansion in terms of $z^a$.

To derive an explicit expression, we separate the sums over $r$ in (\ref{logG}) into even ($r=2s$) and odd ($r=2s+1$) terms. After some manipulations we obtain the form
\be
\ln G_{k,mn} (\za,q) = (n-m) \ln z + {n^2 - m^2 \over 2} \ln q
 + \sum_{a=1}^{\infty} z^a P_{k,mn;a} (q ;\omega)
\label{logGG}\ee
\vskip -0.5cm
\noindent
with
\bea
&&P_{k,mn;a} (q ;\omega) =\bb\bb\bb \sum_{l_1, l_2, \ldots, l_{j} \atop { \rm{compositions}}\;{\rm of}\;a} 
\bb\bb\bb\bb c_2 (l_1,l_2,\ldots,l_{j} ) \,\,
q^{\half\scaleobj{1.2}{\sum_{i=1}^j {(i-1)} l_i} \,-\half \scaleobj{1.2}{  S(l_2\hskip -0.03cm , l_4,\dots \hskip -0.02cm )}}  \, \times \label{Papa}\\
&&\left(
\omega^{{\scaleobj{1.2}{a -2 S(l_2\hskip -0.03cm , l_4,\dots \hskip -0.02cm )}}}
\bb
\sum_{s = \max(\lfloor{m+1-{j\over 2}}\rfloor,0)}^{\min(\lfloor{k+1-{j\over 2}}\rfloor,n)} 
\bb q^{\scaleobj{1.2}{sa}}
+
\omega^{{\scaleobj{1.2}{-a + 2  S(l_2\hskip -0.03cm , l_4,\dots \hskip -0.02cm )}}}\,
q^{\scaleobj{1.2}{  S(l_2\hskip -0.03cm , l_4,\dots \hskip -0.02cm )}} 
 \sum_{s = \max(\lfloor{m+{1-j\over 2}}\hskip -0.03cm\rfloor,0)}^{\min(\lfloor{k+{1-j\over 2}}\rfloor,n)}
 \bb q^{\scaleobj{1.2}{sa}} \right) \nonumber
\eea
where $\lfloor \hskip 0.02cm \cdot \hskip 0.02cm \rfloor$ is the `floor' (integer part) function and $S(l_2,l_4,\hskip -0.03cm\dots\hskip -0.02cm )$
is the sum of even-order $l_i$
\be
S(l_2,l_4,\dots) =\sum_{i=1}^{\lfloor{j\over 2}\rfloor}l_{2i}
\ee
$P_{k,mn;a} (q ;\omega)$ is necessarily real, although this is rather obscured in the expression (\ref{Papa}): for $\omega$ complex,
each term in (\ref{Papa})
is complex, and terms from different compositions $l_1,\dots ,l_j$ combine to a real. The actual form of
$c_2 (l_1,\dots,l_j )$ is needed to achieve reality.

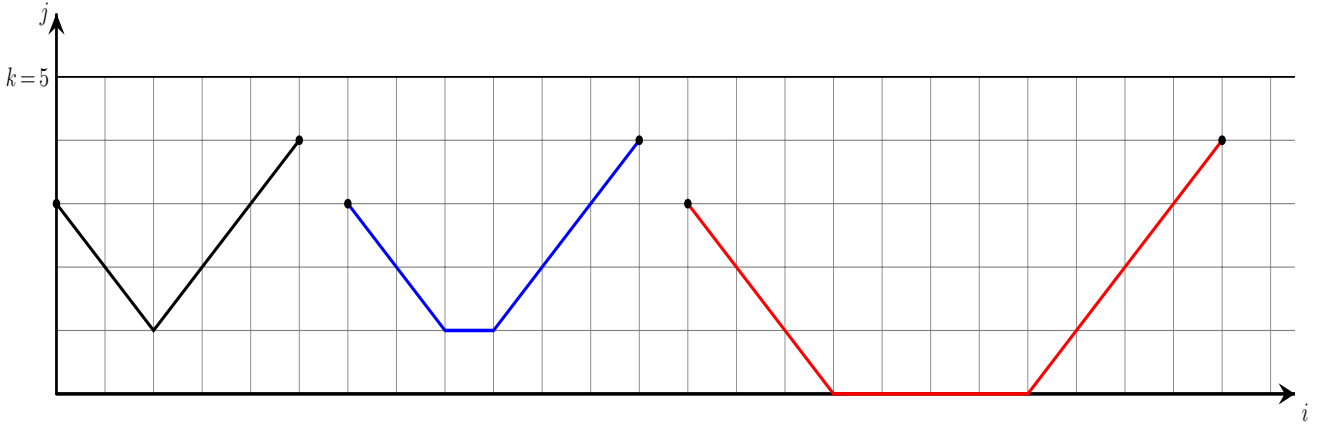
\begin{figure}\vskip -1cm
\hskip -0.9cm
\resizebox{17.5cm}{5.8cm}
{\begin{tikzpicture}

\draw[help lines, gray] (0,-0.02) grid (25.5,5.01);

\tikzset{big arrow/.style={decoration={markings,mark=at position 1 with {\arrow[scale=3,#1,>=stealth]{>}}},postaction={decorate},},big arrow/.default=black}

\draw[ultra thick,-] (0,0) -- (25.5,0);
\draw[ultra thick,-] (0,-0.02) -- (0,6);
\draw[thick,-] (0,5) node[left] {$k\bb=\bb 5$} -- (25.5,5);


\draw[big arrow] (0,0) -- (25.5,0) node[below right] {$i$};

\draw[big arrow] (0,-0.02) -- (0,6) node[left] {$j$};

\draw[ultra thick,black,-](0,3)--(2,1)--(5,4); 
\draw[ultra thick,blue,-](6,3)--(8,1)--(9,1)--(12,4); 
\draw[ultra thick,red,-](13,3)--(16,0)--(20,0)--(24,4);

\fill (0,3) circle (2.2pt); \fill (5,4) circle (2.2pt);
\fill (6,3) circle (2.2pt); \fill (12,4) circle (2.2pt);
\fill (13,3) circle (2.2pt); \fill (24,4) circle (2.2pt);

\end{tikzpicture}}

\vskip 0cm
\caption{\small{Minimal area Motzkin meanders of various lengths from $m=3$ to $n=4$: the first one (black) of length $l=5$
is an unrestricted ``gorge'', while the second one (blue) of length $l=6$ necessitates a horizontal step at the bottom. Their area of 11.5 and 12.5 is given by the top expression in (\ref{minA}). The third one (red) of length $l=11$ is a ``valley'' restricted by the floor. Its area of 12.5 is given by the bottom expression in (\ref{minA}). Paths of maximal area restricted
by the ceiling have similar but inverted shapes.}}\label{Minimal area}
\end{figure}

$P_{k,mn;a} (q ;\omega)$ is a polynomial in $q$ since the fractional powers of $q$
combine to an integer:
\be
\half \sum_{i=1}^j (i-1) l_i - \half S(l_2,l_4,\dots) = l_3 + l_4 + 2l_5 + 2l_6 + \dots
\ee
An examination of the terms in (\ref{Papa}) determines its maximal power in $q$ (degree) as
\be \hskip -1cm
{\text {max~power~in~$q$~of}}~P_{k,mn;a} (q) = \left\{ \begin{matrix}
a n +\scaleobj{1.2}{\lfloor {a^2 \over 4} \rfloor}~~ & ,~~ a \le 2k-2n \\
 & \\
ak-(k-n)^2~& ,~~a > 2k-2n\\
\end{matrix} \right.
\label{degP}\ee
The minimal power in $q$ (with nonzero coefficient) can also be extracted:
\be \hskip -1cm
{\text {min~power~in~$q$~of}}~P_{k,mn;a} (q) = \left\{ \begin{matrix}
a m -\scaleobj{1.2}{\lfloor {a^2 \over 4} \rfloor}~~ & ,~~ a \le 2m \\
 & \\
m^2~& ,~~a > 2m\\
\end{matrix} \right.
\label{minP}\ee
The above results have clear geometric interpretations (see fig.~\ref{Minimal area}):
the prefactors in the expression (\ref{typara}) for
$G_{k,mn} (z,\omega,q)$, leading to the log terms in (\ref{logGG}), correspond to the minimal length
$n-m$ that a path connecting points at heights $m$ and $n$ can have, and the minimal area $n^2/2 - m^2/2$
that such a straight path will have. Terms $z^a$ correspond to an additional length $a$ over the minimal one,
and the exponential $\exp P_{k,mn;a} (q ;\omega)$ accounts for the additional area of such nonminimal paths.

Since the degree of $P_{k,mn;a} (q ;\omega)$ (\ref{degP}) is a convex function of $a$, the degree of the
corresponding terms of order $z^a$ in $\exp P_{k,mn;a} (q ;\omega)$ is the same as that of $P_{k,mn;a} (q ;\omega)$.
Therefore, (\ref{degP}) gives the maximal excess area of a path of length $n-m +a$: the upper expression corresponds
to a ``roof'' path that cannot touch the ceiling, for which the height restriction is irrelevant, while the lower expression
corresponds to a ``flattened roof'' path that grazes the ceiling.

Similarly, since the lowest power of $P_{k,mn;a} (q ;\omega)$ in (\ref{minP}) is a concave function of $a$,
the minimal power in $q$ of the term of order $z^a$ in $\exp P_{k,mn;a} (q ;\omega)$ is the same as that of
$P_{k,mn;a} (q ;\omega)$. Therefore, (\ref{minP}) gives the minimal excess area of a path of length $n-m +a$:
the upper expression corresponds to a ``gorge'' path that cannot touch the floor, while the lower expression
corresponds to a ``valley'' path that creeps on the floor (see fig.\ref{Minimal area}).

The above expressions depend only on $n$ (for maximal area) or $m$ (for minimal area). However, the expressions
of the total maximal or
minimal area in terms of the total length $l = n-m+a$ become symmetric in $n,m$:
\be \hskip -0cm
A_{\max} = \left\{ \begin{matrix}
\scaleobj{1.2}{\lfloor \left( {m+n+l \over 2}\right)^2 \bb\rfloor} -\scaleobj{1.2}{ {m^2 + n^2 \over 2}}~~ & ,~~ l+m+n \le 2k \\
 & \\
k(m+n+l-k)-\scaleobj{1.2}{{m^2 + n^2 \over 2}}~& ,~~ l+m+n > 2k\\
\end{matrix} \right.
\label{maxA}\ee
\be \hskip -0cm
A_{\min} = \left\{ \begin{matrix}
\scaleobj{1.2}{ {m^2 + n^2 \over 2}}- \scaleobj{1.2}{\lfloor \left( {m+n-l \over 2}\right)^2\bb \rfloor} ~~ & ,~~ l \le m+n \\
 & \\
\scaleobj{1.2}{{m^2 + n^2 \over 2}}~& ,~~ l >m+n \\
\end{matrix} \right.
\label{minA}\ee
It can be checked that the above reproduce the actual areas of the maximal and minimal area paths.
When $m+n+l$ is odd, in which case the top or bottom of unrestricted paths acquires a single horizontal link,
the floor function adds a correction of $\pm {1\over 4}$ corresponding to the truncated triangular part.

\section{Conclusions and discussion}

The use of a Hamiltonian framework in combination with the two-step construction, exclusion statistics and bosonization
allowed for the derivation of generating functions for Motzkin paths including dual variables for length, area, as well
as various top and bottom ``events''. The relation to quantum statistics of exclusion $g=2$ identified in~\cite{DyckPO}
for Dyck paths persists in the case of Motzkin paths, allowing the calculation of these functions and providing cluster decomposition techniques used to
derive alternative expressions for the logarithm of the generating functions in terms of sums of compositions.

It is interesting to compare our results with other work done in the literature and put them in the proper context.
The generating functions studied in this paper are essentially embedded in the general class of Motzkin polynomials,
defined as sums over paths with height- and step-specific weights \cite{Flaj,OstJeu}. Specifically, if each up, horizontal,
or down step ending at height $j$ is assigned a variable $u_j$, $h_j$ and $d_j\bb=\bb 1$, respectively, and walks are weighted by
the product
of such variables, then (using a notation consistent with our conventions) $P_{l|k}^{(m,n)} (\{u_j\},\{h_j\})$ is a
polynomial in the variables $h_0, h_j , u_j$, $j = 1,2,3,\dots$ arising from summing the weights of all meanders of length
$l$ starting at $m$ and ending at $n$ with a ceiling $k$, while $P_l (\{u_j\},\{h_j\})$ is the corresponding sum
over unrestricted excursions. It should be clear that $G_{\infty,00} = G$ is related to $P_l$ as
\be
G (\za,q) = \sum_{l=0}^\infty P_l (\{u_j\},\{h_j\}) ~,~~\text{with}~
u_j = z^2 q^{2j-1} ~,~ h_j = z_h q^{j+1/2}
\ee
(An analogous relation between $G_{k,mn}$ and $P_{l|k}^{(m,n)}$ fails, since the convention $d_j\bb=\bb 1$ only
considers the area of up-links, which is not related to that of down-links if $m\neq n$.) In fact, an explicit
combinatorial expression for the polynomial $P_l$ involving multiple sums was given in \cite{Flaj,OstJeu}.
Adapting it to our case and notation it implies
\be
G(\za,q) = \sum_{\{m_j , n_j \ge 0\}} {1\over (1-z_h)^{n_1+1}}\prod_{i=1}^\infty {m_i+n_{i+1}+n_i-1 \choose m_i ~~,~~ n_{i+1}}
\left(z^2 q^{2i-1} \right)^{n_i} \left( z_h q^i \right)^{m_i}
\label{MoPoly}\ee
with the usual definition of the choose-symbol
\be
{M \choose N ,K} = {M! \over N!\, K!\, (M\bb-\bb N\bb-\bb K)!}
\ee
(The restriction $\sum_i 2n_i + \sum_j m_j = l$ in \cite{OstJeu} is relaxed since we sum over all $l$. We also rearranged the factors and performed the sum over $m_0$). This is a compact-looking expression, but in fact it involves infinitely many sums.
(We also have no expression for the more general $G_{k,mn}$.)  Our formulae
(\ref{wow1}) and (\ref{wow2}), on the other hand, involve only double sums. 
We can, actually, evaluate the sums in (\ref{MoPoly})
sequentially, in the order $m_1,n_1,m_2,n_2\dots$, but at the end we obtain an infinitely
nested expression, essentially the Rogers-Ramanujan-like continued fraction that will be given in the sequel.

Generating functions counting the number of horizontal steps $l_h$ were considered in \cite{Sula} by including
a weight $t^{l_j}$,
but not the remaining steps nor the area. Interestingly, in the same paper the first and second moments of the area
under the paths were considered and shown to satisfy specific recursion relations. These would correspond to the
$O(\epsilon)$ and $O(\epsilon^2 )$ terms in an $\epsilon$ expansion of our 
$G (z\bb=\bb 1,z_h\bb =\bb1,q\bb=\bb e^\epsilon)$. Quantities
related to path area were also considered in \cite{JRW}.

The work whose results come closest to ours is \cite{RSOS}, which studied Motzkin excursions (floor-to-floor paths), also
counting touch-ups and touch-downs (although nor creep-us and creep-downs), and calculated the length-area generating function.
The results were obtained by combinatorially driving recursion relations analogous to (\ref{concur}) and solving them using
various ingenious mathematical techniques. The resulting generating function is then the ratio of quantities expressed
in terms of generalized hypergeometric functions. These results should match our results (\ref{Gexcur}), (\ref{hag})
(with $m=n=0$ and thus $G_{m-1} =0$, and $s=\s =1$) and, in particular, (\ref{wow1}) and (\ref{wow2}). However,
the expressions in (\ref{wow1},\ref{wow2}) look quite different than the ones in \cite{RSOS}, involving sums of Chebyshev
polynomials rather than generalized hypergeometric functions. Such differences of form are expected, given the nontrivial
nature of the results and the various identities that could be used in reshaping them, and already \cite{RSOS} noted
that their results for the half-plane ($k=\infty$, no ceiling) and the slit ($k$ finite) looked `dramatically different'.
The reconciliation between our results and those of \cite{RSOS} is an open interesting mathematical task.

The generating functions derived in the present work satisfy several recursion relations, as stated in section 3.3. Such relations
for Motzkin polynomials were derived in several papers and invariably lead to expressions related to the Rogers-Ramanujan
continued fraction.
Specifically, the recursion relation (\ref{concur}) for $G_k (z,q)$ can be iterated leading to a continued fraction. To put it in a clean form, we define
\be
\lambda = \omega+\omega^{-1} = {z_h \over z}~,~~w=z^{-1} ~,~~ u = q^{-1}~,~~
g_k (w) = z G_k (z,\omega,q)
\ee
In this parametrization, the recursion relation becomes
\be
g_k (w) = {1\over w-\lambda -g_{k-1} (uw)}
\ee
Iterating this relation with the final condition $G_{-1} (z,q) = 0$ we obtain\vskip -0.7cm
\be
g_k (w) = {1 \over w\bb-\bb\lambda - {{\scaleobj{1.2}1} \over {\scaleobj{1.2}{wu-\lambda-} \scaleobj{1.2} {1 \over \cdots {{1 \over wu^k - \lambda}}}}}}
\ee
This is a truncated version of a continued fraction related to the Rogers-Ramanujan's identity. Other forms, involving
directly $G_k$, are readily obtainable.

We conclude with  some possible directions for future research. The most interesting and relevant next task would be
to use the results in this work to derive physical properties of statistical systems described in terms of Motzkin paths.
Several systems can be mapped to such paths, the canonical one being linear polymers on the plane in a slit
(represented by the space between the floor and ceiling of the paths), with potential adsorbing interactions whenever
the polymer bounces off or sticks to the two boundaries (our $t,s,\T,\s$ terms in section \ref{tobo}). As a physical model,
weighting by the area underneath the path accounts for polymers with a different solvent from the area above the path by
assigning an energy depending on the area, $q$ playing the role of the corresponding Boltzmann factor. The area-length
generating function then reproduces the
partition function of the polymer, and the free energy of the model can be determined from that generating function
and would determine critical transition properties of the model (as in \cite{Ren} for $q=1$). Such critical properties
for $q \neq 1$ have not been explored, to the best of our knowledge, and constitute an important open problem and
obvious next step .

On the mathematical side, the results in this work most likely admit further refinement and elaboration. For instance,
the explicit expressions
for $F_k$, $G_{k,mn}$ and $\tG_{k,mn}$ presented in this paper are not unique, as they satisfy several identities
and recursion relations, and alternative forms are possible, as we already noted in the comparison with the results
in \cite{RSOS}. In addition, it would be desirable to have
expressions for the cluster coefficients  involving directly $z,z_h$ as opposed to $z,\omega$, as in (\ref{wow2}) for the
partition function, making the dependence on $z_h$ clearer. Such a rewriting remains to be achieved.

The enumeration of Motzkin paths according to their length can be expressed in terms of trinomial coefficients
in the expansion of $(x + 1 + x^{-1} )^l$, the term $x^n$ identifying the (unrestricted from floor or ceiling)
Motzkin paths of total climb $n$ (or descent, if $n<0$). This method was used in \cite{Ren} to find the length
generating function ($q=1$) of unrestricted Motzkin paths ($k=\infty$), and derive critical properties of the corresponding
statistical mechanical model. The use of this
technique, however, becomes cumbersome when dealing with restricted paths ($k<\infty$) and fails to address
the more interesting case of length {\it and} area generating functions ($q \neq 1$). In a related development,
the expressions of length generating functions ($q=1$) for restricted Dyck, Motzkin and more general paths with a number of
possible up and down steps, and arbitrary weights associated to each kind of step, have been related to skew-Schur
functions \cite{KP}. Including the area counting variable $q$ would generalize these generating functions to $q$-deformed
versions of skew-Schur functions, both in the case of Dyck paths and for Motzkin paths. This points to a possible
generalization of the trinomial method involving $q$-deformed polynomial expressions. Such a generalization and the
related skew-Schur functions and their properties remain an interesting topic for further mathematical study.

The $q\bb =\bb 1$ limit of Dyck or Motzkin paths is intimately related to compositions of a large number of $SU(2)$ spins:
spin-$\half$ individual spins are related to Dyck paths, while Motzkin paths correspond to spin-$1$.
In \cite{PS}, the combinatorics and statistics of such compositions of general spin-$s$ components were studied using
generating function and partition function techniques, and a corresponding large-$N$ phase transition was identified.
Symmetric (bosonic) and antisymmetric (fermionic) spin compositions were also studied in \cite{PS} and led to novel
statistical properties. The existence of a ceiling at $n=k$ for paths would correspond to deforming the spin group to
the `quantum group' $SU(2)_{_Q}$ with $Q = \exp[2\pi i/(k+1)]$, which has irreducible representations of dimension
up to $k+1$. It would be interesting to further explore the connection between the two systems (spins and paths).
Investigating the physical meaning of weighting the spin compositions with an exponential factor proportional
to the ``area'' of the specific composition channel, as for paths, and, conversely, the concept of symmetric or antisymmetric
weighing of paths, as for fermionic or bosonic spins, and the possibility of a phase transition in the statistics of paths,
are fascinating topics that deserve further exploration.

Finally, there are other generalizations of paths that have been studied in the literature. For instance, ``colored''
Motzkin paths in which each link can come in one of several `colors', and $k$-Motzkin paths in which horizontal
steps are of length $k$ have been considered. Further, paths with more general increments, such as
Lukasiewicz paths, have been studied, and there are other possible generalizations that have not. All such paths can
be treated in the Hamiltonian framework, and their generating functions can 
be related to the secular Hamiltonian of the process, as in sections 2 and 3 of this paper. However, 
explicit expressions for the generating function of such systems would require the evaluation of secular determinants
as in section 4 of this paper, which may not be tractable. It appears that at least a class of such walks can be related
to quantum exclusion statistics, but for exclusion higher than 2 and for more general one-body spectra.
The statistical mechanics of general-$g$ exclusion
systems with an arbitrary discrete energy spectrum have recently been derived using techniques closely related to the
ones in the present work \cite{OP}. Using these techniques, the generating functions and statistics of generalized paths
could be derived. We defer a full treatment of these cases to a future publication.

\vskip 0.5cm

\noindent
{\bf Acknowledgements}

\noindent
This work was supported in part by grant NSF-PHY-2112729 and a PSC-CUNY grant. I would also like to thank Thomas Prellberg and the two anonymous referees for constructive comments that helped improve the manuscript.

\vskip 0.3cm

%

\end{document}